\documentclass[a4paper,11pt]{article}
\pdfoutput=1 
\usepackage{jcappub} 

\usepackage{amssymb}
\usepackage{mathtools}
\usepackage{amsmath}
\usepackage{amstext}
\usepackage{graphicx,epsfig}
\usepackage{epsfig}
\usepackage{verbatim} 	
\usepackage{caption}
\usepackage{fancybox}
\usepackage{slashed}
\usepackage{afterpage}
\usepackage{color}
\usepackage{ulem}
\usepackage{enumitem}
\usepackage{subfigure}
\usepackage{parskip}
\usepackage{dsfont}
\usepackage{tabu}
\usepackage{tikz}
\usepackage{booktabs}
\usepackage[numbers,sort&compress]{natbib}
\usepackage{cancel}

\usepackage{bbding}
\usepackage{wasysym}
\usepackage{multirow,hhline,array}
\usepackage[utf8]{inputenc}

\definecolor{darkred}{rgb}{0.7,0.3,0.3}
\definecolor{darkgreen}{rgb}{0.2,0.7,0.3}
\definecolor{greyish}{rgb}{.90,.90,.90}
\definecolor{greyish2}{rgb}{.96,.96,.96}
\definecolor{darkblue2}{rgb}{0.3,0.4,0.9}
\usepackage{pifont}
\usepackage{xcolor,colortbl}
\usepackage{tcolorbox}

\definecolor{navyblue}{rgb}{0.0, 0.0, 0.5}
\definecolor{royalblue}{rgb}{0.25, 0.41, 0.88}
\definecolor{cadmiumgreen}{rgb}{0.0, 0.42, 0.24}
\definecolor{blue-violet}{rgb}{0.54, 0.17, 0.89}
\definecolor{darkviolet}{rgb}{0.58, 0.0, 0.83}
\definecolor{orange(colorwheel)}{rgb}{1.0, 0.5, 0.0}

\usepackage{hyperref}
\hypersetup{colorlinks=true,linkcolor=royalblue,citecolor=magenta,pdfencoding=auto}
\usepackage{pifont}
\usepackage{dcolumn}
\usepackage{colortbl}

\newcommand{\D}{{\rm d}}

\newcommand{\overbar}[1]{\mkern 1.5mu\overline{\mkern-1.5mu#1\mkern-1.5mu}\mkern 1.5mu}

\renewcommand\[{\left[}

\newcommand{\Mpl}{M_{\text{Pl}}}

\definecolor{magenta(process)}{rgb}{1.0, 0.0, 0.56}

\definecolor{darkspringgreen}{rgb}{0.09, 0.45, 0.27}

\definecolor{royalblue(web)}{rgb}{0.25, 0.41, 0.88}

\def\leq{\raise 0.4ex\hbox{$<$}\kern -0.8em\lower 0.62ex\hbox{$-$}}
\def\geq{\raise 0.4ex\hbox{$>$}\kern -0.7em\lower 0.62ex\hbox{$-$}}
\def\lsim{\raise 0.4ex\hbox{$<$}\kern -0.8em\lower 0.62ex\hbox{$\sim$}}
\def\gsim{\raise 0.4ex\hbox{$>$}\kern -0.7em\lower 0.62ex\hbox{$\sim$}}

\title{\centering Black Hole Hair \\ from Scalar Dark Matter}

\author[a]{Lam Hui,}
\author[b]{Daniel Kabat,}
\author[a]{Xinyu Li,}
\author[a]{Luca Santoni,}
\author[a]{Sam S. C. Wong}

\affiliation[a]{Department of Physics, Center for Theoretical Physics, Columbia University, 538 West 120th Street, New York, NY 10027, U.S.A.}
\affiliation[b]{Department of Physics and Astronomy, Lehman College, City University of New York, Bronx NY 10468, USA}

\emailAdd{lh399@columbia.edu}
\emailAdd{daniel.kabat@lehman.cuny.edu}
\emailAdd{xinyu.li@columbia.edu}
\emailAdd{ls3598@columbia.edu}
\emailAdd{sw3266@columbia.edu}

\abstract{\noindent
We show that a black hole surrounded by scalar dark matter develops
scalar hair. This is the generalization of a phenomenon pointed out
by Jacobson \cite{Jacobson:1999vr}, that a minimally coupled scalar with a non-trivial
time dependence far away from the black hole would endow the black
hole with hair. In our case, the time dependence arises from the
oscillation of a scalar field with a non-zero mass. We systematically
explore the scalar profile around the black hole for different
scalar masses. In the small mass
limit, the scalar field has a $1/r$ component at large radius $r$, consistent with
Jacobson's result. In the large mass limit (with the Compton
wavelength of order of the horizon or smaller), the scalar field has a
$1/r^{3/4}$ profile yielding a pile-up close to the horizon, while
distinctive nodes occur for intermediate masses.
Thus, the dark matter profile around a black hole, while challenging
to measure, contains information about the dark matter
particle mass. As an application, we
consider the case of the supermassive black hole at the center of M87,
recently imaged by the Event Horizon Telescope. Its horizon size
is roughly the Compton wavelength of a scalar particle of mass
$10^{-20}$ eV. We consider the implications of the expected scalar
pile-up close to the horizon, for fuzzy dark matter at a mass of
$10^{-20}$ eV or below.
}

\begin{document}
\maketitle
\flushbottom

\section{Introduction}
\label{sec:intro}
Like all no-go theorems, the well known no-scalar-hair theorem of Bekenstein \cite{Bekenstein:1971hc}
can be violated---the theorem is correct of course, but its assumptions
can be circumvented.
Among the assumptions that go into the theorem, an important one is that the
scalar field vanishes far away from the black hole. Jacobson \cite{Jacobson:1999vr} pointed out that, in the case of a 
massless, minimally coupled scalar, giving the scalar field far away a linear time-dependence 
is sufficient to generate hair for the black hole.\footnote{For alternative ways to circumvent Bekenstein’s theorem, see for example
\cite{Sotiriou:2014pfa,Babichev:2016fbg,Doneva:2017bvd,Silva:2017uqg,Antoniou:2017acq,Antoniou:2017hxj,Macedo:2019sem} and the review \cite{Herdeiro:2015waa}. For an extension of the no-scalar-hair 
theorem to the galileon, see \cite{Hui:2012qt}.} In other words, in a Schwarzschild background,\footnote{Jacobson also derived the analogous result in a Kerr background. We focused on the non-rotating case in this paper.}
\begin{eqnarray}
\D s^2 = - \left( 1 - {r_s \over r} \right) \D t^2 + \left(1 - {r_s \over r}\right)^{-1} \D r^2 + r^2 \D \theta^2
+ r^2 {\rm sin\,}\theta {}^2 \D \phi^2 \, ,
\label{static}
\end{eqnarray}
where $r_s \equiv 2 G M_{\rm BH}$ is the Schwarzschild radius
($M_{\rm BH}$ being the black hole mass and $G$ being the Newton
constant),  Jacobson showed that the equation
$\Box \phi = 0$ has, in addition to the trivial solution $\phi = 0$ (which would be
consistent with Bekenstein's theorem), a solution of the following form for the
scalar $\phi$:
\begin{eqnarray}
\phi \propto \left( t + r_s {\,\rm log\,} (1 - {r_s \over r}) \right) \, .
\label{jacsol}
\end{eqnarray}
At large $r$, this asymptotes to $t$ plus a $- r_s^2 / r$ tail.
In other words, $\phi$ does not vanish at spatial infinity but rather takes on a linear 
time dependence. The coefficient of the $1/r$ tail can be interpreted as the scalar charge
of the black hole.
A non-trivial aspect of this solution is that it is regular at the horizon,
which is easiest to see by noting that at the horizon $t + r_*$ 
(often called $v$, the Eddington-Finkelstein time) is regular,
where $r_*$ is the tortoise coordinate $r_* \equiv r + r_s {\,\rm log\,}(r/r_s - 1)$. 
The scalar field $\phi$ is finite at the horizon, and so is $\partial_\alpha \phi \partial^\alpha \phi$.

The coefficient of the $1/r$ tail is often identified as the scalar
charge of the black hole, much like the coefficient of the $1/r$ tail
in the gravitational potential is identified as the mass of the black
hole.\footnote{To be more precise, one could for instance define
the scalar charge to be $Q \equiv {\cal C} r_s^2 \Mpl$, where ${\cal
  C}$ is the proportionality constant in $\phi = {\cal C} (t + r_s {\rm
  \,log\,} [1 - r_s/r] )$, and $\Mpl = 1/G$. In this way, $Q$ has mass dimension. We will
not need this (somewhat arbitrary) definition in the rest of paper.}
In this paper, we take a more general view of what constitutes the scalar hair of
a black hole: it needs not have a $1/r$ spatial profile; any
non-trivial spatial profile around the black hole is potentially
interesting from an observational point of view.

Jacobson's insight is that a black hole can be endowed with a scalar charge (or scalar hair)
by imposing the boundary condition that the scalar has a non-zero time derivative far away from the black hole. The original motivation of Jacobson was to apply this to a cosmologically 
evolving scalar, in which case the proportionality constant in
Eq. (\ref{jacsol}) is set by the Hubble expansion rate $H$. Given the
vast disparity in scale between $r_s$ and $1/H$, this can be
interpreted as a very small scalar charge (or very small charge to
black hole mass ratio).

This raises an obvious question: how about cases where the time derivative
far away from the black hole is much larger? 
A natural setting for this is a scalar with a non-zero mass, which thus oscillates in time. 
An appealing scenario is one where dark matter is comprised of such a scalar,
which inevitably surrounds the black hole.
The question we wish to address in this paper is:
{\it what scalar profile should we expect 
around a black hole embedded within a dark matter halo made out of a scalar field with non-vanishing mass?}
A natural candidate for scalar dark matter is an axion or axion-like-particle.
Possible masses range from $10^{-22}$ eV to $10^{-3}$ eV
\cite{Baldeschi:1983mq,Turner:1983he,Press:1989id,Sin:1992bg,Matos:1992qx,Peebles:2000yy,Goodman:2000tg,
  Hu:2000ke,Lesgourgues:2002hk,Amendola:2005ad,Svrcek:2006yi,Arvanitaki:2009fg,Schive:2014dra,Hui:2016ltb}. 
The QCD axion tends to occupy the higher mass range, while axions in string theory
can span the whole range. At the lowest mass end is what is sometimes
referred as fuzzy dark matter \cite{Hu:2000ke,Schive:2014dra,Hlozek:2014lca,Mocz:2015sda,Hui:2016ltb,Nori:2018hud,Veltmaat:2018dfz}.
Our goal in this paper is the work out the scalar profile for the full range of
possible scalar masses.

Addressing the question of interest requires revisiting the massive Klein-Gordon equation in a Schwarzschild background. 
It is not surprising there is a large literature on this subject. 
For instance, the solution to the Klein-Gordon equation, in certain limiting cases
such as large or small radius, was given by
Unruh \cite{Unruh:1976fm} and Detweiler \cite{Detweiler:1980uk}.
The former focused on computing the absorption cross section of the scalar by the black hole,
while the latter emphasized the instability associated with super-radiance.
More recently, it was pointed out that the exact solutions of the
Regge-Wheeler equation and  Klein-Gordon equation in Schwarzschild
space-time is a special function known as the confluent Heun function
\cite{Fiziev:2005ki,Fiziev:2006tx,Bezerra:2013iha,Vieira:2014waa}. 
A number of authors \cite{Konoplya:2006br,Barranco:2012qs} used the confluent Heun function to 
compute the quasi-normal spectrum, highlighting in particular the
long-lived modes. Related to our discussion are papers on 
the effect of dark matter on binary inspiral
(e.g. \cite{Macedo:2013qea}), and the Jacobson effect due to  a black
hole moving in an inhomogeneous scalar background, pointed out by
Horbatsch and Burgess \cite{Horbatsch:2011ye}.
Building on the prior work, our goal in this paper is a rather modest one:
we use the exact Heun solution to explore the full range of masses,
and frame the discussion in terms of scalar hair \`a la Jacobson.

It is also worth mentioning there is a large literature on
super-radiance around black holes \cite{1972JETP351085Z,1973JETP3728S,Arvanitaki:2009fg,Herdeiro:2014goa}. 
The super-radiance cloud can be quantum
mechanically generated (when the relevant Compton wavelength is around
the horizon size of a rotating black hole), and needs not be related to
dark matter. 

To set the stage we discuss the scales relevant to our problem.
First, there
is the Schwarzschild radius $r_s$:
\begin{equation}
r_s = 2.95 {\,\rm km} \left( {M_{\rm BH} \over M_\odot} \right)
\, .
\end{equation}
Interesting values for $r_s$ range from $\sim 30$ km for typical
LIGO black holes \cite{Abbott:2016blz}, to $\sim 10^7$ km for LISA
black holes \cite{AmaroSeoane:2012je,AmaroSeoane:2012km}, 
to $10^{10}$ km for pulsar timing array PTA black holes \cite{Hobbs:2009yy,Gentile:2018nrq}.\footnote{All experiments LIGO, LISA and PTA are of course sensitive to a range
  of black hole masses. See \cite{Abbott:2016blz,AmaroSeoane:2012je,AmaroSeoane:2012km,Hobbs:2009yy,Gentile:2018nrq} for details.}
The scales of $10^7$ km and $10^{10}$ km are also
roughly the size of the black hole at the center of the Milky Way and
that in M87, relevant for the Event Horizon Telescope EHT \cite{Akiyama:2019cqa}. 

We are interested in how this scale compares to the Compton wavelength
$1/m$.\footnote{The Compton wavelength is $1/m = 1.97 {\,\rm km\,} \left( {10^{-10} {\,\rm eV} / m}
  \right)$. Note that the Planck constant $\hbar$ and the speed of
  light $c$ are set to unity by default.} Alternatively, we compare the scalar mass $m$ against
$1/r_s$, expressed in eV:
\begin{eqnarray}
r_s^{-1} = 6.7 \times 10^{-11} {\,\rm eV} \left( {M_\odot \over M_{\rm
  BH}} \right) \, .
  \end{eqnarray}
Thus, $1/r_s$ ranges from $\sim 10^{-11}$ eV for LIGO black holes
to $\sim 10^{-17}$ eV for LISA black holes to $\sim 10^{-20}$ eV for
PTA black holes.

While the background geometry is well described by the Schwarzschild
metric close enough to the black hole, sufficiently far from it the
gravitational influence of the surrounding matter is non-negligible.
We define a scale $r_i$---we call it the radius of sphere of impact---as the
radius within which the black hole dominates the geometry.\footnote{In the literature, the term radius of sphere of influence is
  often used to describe the radius within which the black hole's mass
  dominates over the mass of the enclosed, surrounding matter.
Our $r_i$ coincides with that definition, if $v_{\rm typical}$
is chosen to be the dynamical velocity associated with enclosed matter mass.
}
In other words, we define $r_s / r_i \equiv v_{\rm typical}^2$ ($c=1$), where
$v_{\rm typical}$ is the typical velocity dispersion of the
surrounding matter (motivated by virial theorem) i.e.\footnote{In this paper, we by default set the speed of light to
  unity, and thus $v_{\rm typical}$ should strictly speaking be
  dimensionless. We restore dimension to $v_{\rm typical}$ to ease
  comparisons with typical astrophysical velocities.}
\begin{eqnarray}
\label{ridef}
{r_s \over r_i} = 10^{-6} \left( {v_{\rm typical} \over 300 {\,\rm km/s}}
  \right)^2 \, .
\end{eqnarray}
In realistic settings, $r_s / r_i$ can range from $\sim 10^{-8}$ to $\sim
10^{-5}$. We assume the backreaction of the scalar on the Schwarzschild
geometry is negligible for $r < r_i$. We will check below that this is
a self-consistent assumption, once we work out the scalar profile.

\begin{figure}
\center 
\includegraphics[scale=0.65]{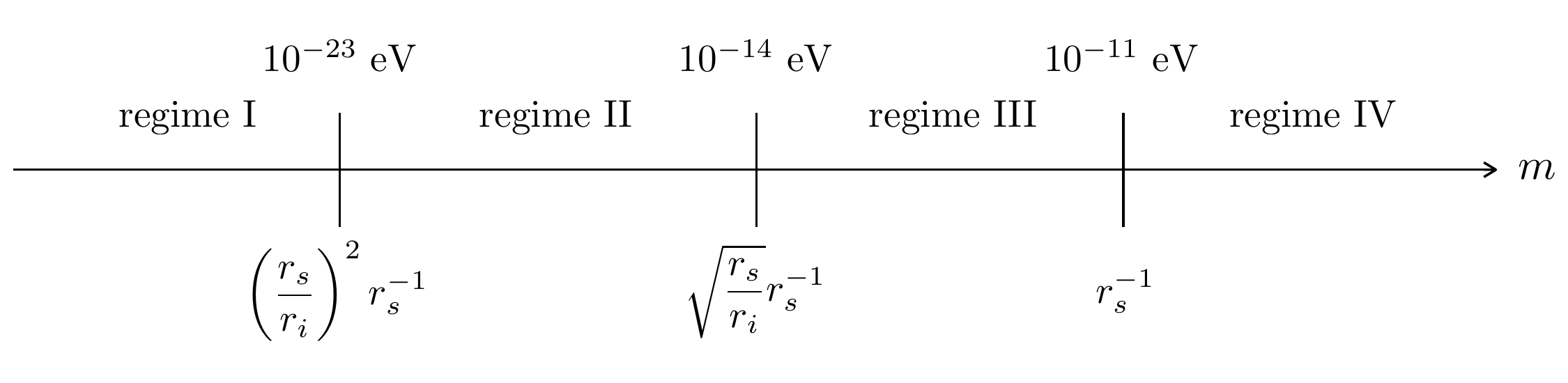}
\caption{The three mass scales that delineate different regimes in the scalar profile. The ratio $r_s/r_i$ is assumed to be $10^{-6}$. The black hole mass is assumed to be about $10 M_\odot$, relevant for typical LIGO events. Multiply all numbers by about a factor of $10^{-6}$ for typical LISA black holes (or for the black hole at the
center of the Milky Way); multiply all by a factor of $\sim 10^{-9}$
for typical PTA black holes (or for the black hole at the center of M87). }
\label{Fig:1}
\end{figure}

Our main goal in this paper is to understand the dependence of the
scalar profile on the scalar mass $m$. As we will see, there are 4 different regimes,
delineated by 3 different scales: regime I --- $m < r_s^{-1}
(r_s/r_i)^2$, regime II --- $r_s^{-1} (r_s/r_i)^2 < m < r_s^{-1}
\sqrt{r_s/r_i}$, 
regime III --- $ r_s^{-1} \sqrt{r_s/r_i} < m < r_s^{-1}$, and
regime IV --- $r_s^{-1} < m$. 
As we scan the regimes from low to high scalar mass, we go
from the wave limit to the particle limit.
Typical values for these scales are given in Fig. \ref{Fig:1}.

The organization of the paper is as follows.
In Sec. \ref{sec:schwarzschild}, we examine the scalar profile in
in a Schwarzschild background, exploring different relevant limits of
the confluent Heun function.
In Sec. \ref{sec:galaxy}, we introduce a toy model for the metric
at distances outside the radius of sphere of impact $r_i$. This allows
us to connect in a simple way the problem of computing the scalar
profile to the problem of scattering. The results of the scattering
computation are summarized in Sec. \ref{sec:galaxy}.
In this paper, we largely focuses on s-waves i.e. the accreting scalar
having no angular momentum. We discuss in Sec. \ref{sec:conclude} and
App. \ref{app:KG} under what condition this is a good approximation, and which results are
modified or remain the same when angular momentum is included.
We conclude in Sec. \ref{sec:conclude}. As an example, we apply our
results to the supermassive black hole in M87, and consider the
implications for fuzzy dark matter whose Compton wavelength is not
much larger than the horizon of the black hole.
Certain technical details are relegated to the Appendices: 
results for non-zero angular momentum are discussed in Appendix
\ref{app:KG}, asymptotics of the Heun function are worked out in
Appendix
\ref{appendix:asymptotics}, and expressions for the energy-momentum
tensor can be found in \ref{app:flux}.

\noindent \textit{Conventions:} {\it For the rest of the paper, we will set $r_s=
2GM_{\rm BH}=1$ and restore it when needed for the sake of clarity.} 
In other words, whenever the dimension of a quantity does not match
the expected one, the reader can simply put in suitable powers of $r_s$ to
recover the correct dimension.
We denote $\Mpl=G^{-1/2}$ the standard Planck mass.

\noindent \textit{Note added:} As this manuscript was under preparation, a recent paper by 
Wong, Davis and Gregory \cite{Wong:2019yoc} appeared which has some overlap with our work,
in particular regarding the scalar charge in the small mass
regime. Independently, Clough, Ferreira and Lagos \cite{Clough:2019jpm}
explored the same subject we investigate in this paper -- hair
associated with a black hole in an oscillating
scalar background --  using numerical methods that enabled them to include the effects of back-reaction on the solution.

\section{Scalar Profile in the Schwarzschild region ($r<r_i$)} \label{sec:schwarzschild}

We consider a scalar field $\phi$ of mass $m$ in the
Schwarzschild geometry \eqref{static}.
For simplicity we set the Schwarzschild radius $r_s = 1$, though 
it will be restored in a few key expressions.
Our focus will be on spherically-symmetric field configurations.  This is particularly relevant for small scalar masses, as we show in section \ref{sec:conclude} where we discuss the effects
of angular momentum.
A general discussion of the Klein Gordon equation $\left(\nabla_\mu \nabla^\mu -
  m^2\right)\phi = 0$, including angular momentum, can be found in appendix \ref{app:KG}.
  
 Restricting to s-waves the Klein-Gordon equation can be expressed in several different ways. 
For instance:
\begin{eqnarray}
\label{KGeq0}
\left[ -\partial_t^2 - m^2 \left(1 - {1\over r}\right)
+ \left(1 - {1 \over r}\right)^2 \partial_r^2 
+ {2\over r} \left( 1 - {1\over r} \right) \left(1 - {1 \over
  2r}\right) \partial_r \right] \phi = 0 \, ,
\end{eqnarray}
or 
\begin{eqnarray}
\label{KGeq2}
\left[ \partial_t^2 - \partial_{r_*}^2 + m^2 - {m^2 \over r} + {1\over
  r^3} - {1\over r^4} \right] (r \phi) = 0 \, ,
\end{eqnarray}
where $r_*$ is the tortoise coordinate defined by $r_* \equiv r + {\,\rm
  log\,} (r-1)$. The second form of the equation is convenient for
thinking of the problem as a scattering problem, a perspective we will
develop more fully in the next section. 

We seek a spherically symmetric solution $\phi(t,r)$ whose time
dependence is completely captured by $e^{-i\omega t}$:
\begin{equation}
\phi(t,r) \propto e^{-i \omega t} \, .
\label{ansatz}
\end{equation}
The most general solution would involve a superposition of different
frequencies $\omega$, but we will not need that for the problem at hand.
Eq. (\ref{KGeq0}) admits
the following general solution in terms of the confluent Heun function
\cite{Bezerra:2013iha,Vieira:2014waa}:
\begin{eqnarray}
\label{generalsolution}
\phi(t,r) = 
c_1 \, e^{-i \omega t} (r - 1)^{i \omega} e^{i \bar{k} r} \, {\sf
  HeunC}(-2i\bar{k}, 2i\omega, 0, -\omega^2 -\bar{k}^2, \omega^2 +
  \bar{k}^2, 1 - r) \nonumber \\
+ c_2 \, e^{-i \omega t} (r - 1)^{-i \omega} e^{-i \bar{k} r} \, {\sf HeunC}(2i\bar{k}, -2i\omega, 0, -\omega^2 -\bar{k}^2, \omega^2 + \bar{k}^2, 1 - r) \, ,
\end{eqnarray}
where $c_1$ and $c_2$ are constants, and the radial momentum $\bar k$
is defined by:\footnote{Note that $\bar{k}$ is the momentum the particle would have at infinity in the Schwarzschild geometry.  We are denoting it $\bar{k}$ to distinguish it from $k$, the typical momentum the
particle would have within the galaxy, which we will introduce later.
Note also ${\sf HeunC}(\alpha, \beta, \gamma, \delta, \eta, z) =
\exp(-z \alpha) \, {\sf HeunC}(-\alpha, \beta, \gamma, \delta, \eta,
z)$, so the sign of $\bar{k}$ does not matter.
}
\begin{eqnarray}
\omega = + \sqrt{\bar{k}^2 + m^2} \, .
\label{omegak}
\end{eqnarray}
In the non-relativistic limit $\bar{k}$ is related to the energy of the particles by
\begin{equation}
\label{Eeqn}
E = \omega - m = {\bar{k}^2 \over 2 m}
\end{equation}
Note that a bound state $E < 0$ corresponds to imaginary $\bar{k}$.

\begin{figure}
\centerline{\includegraphics[width=6cm]{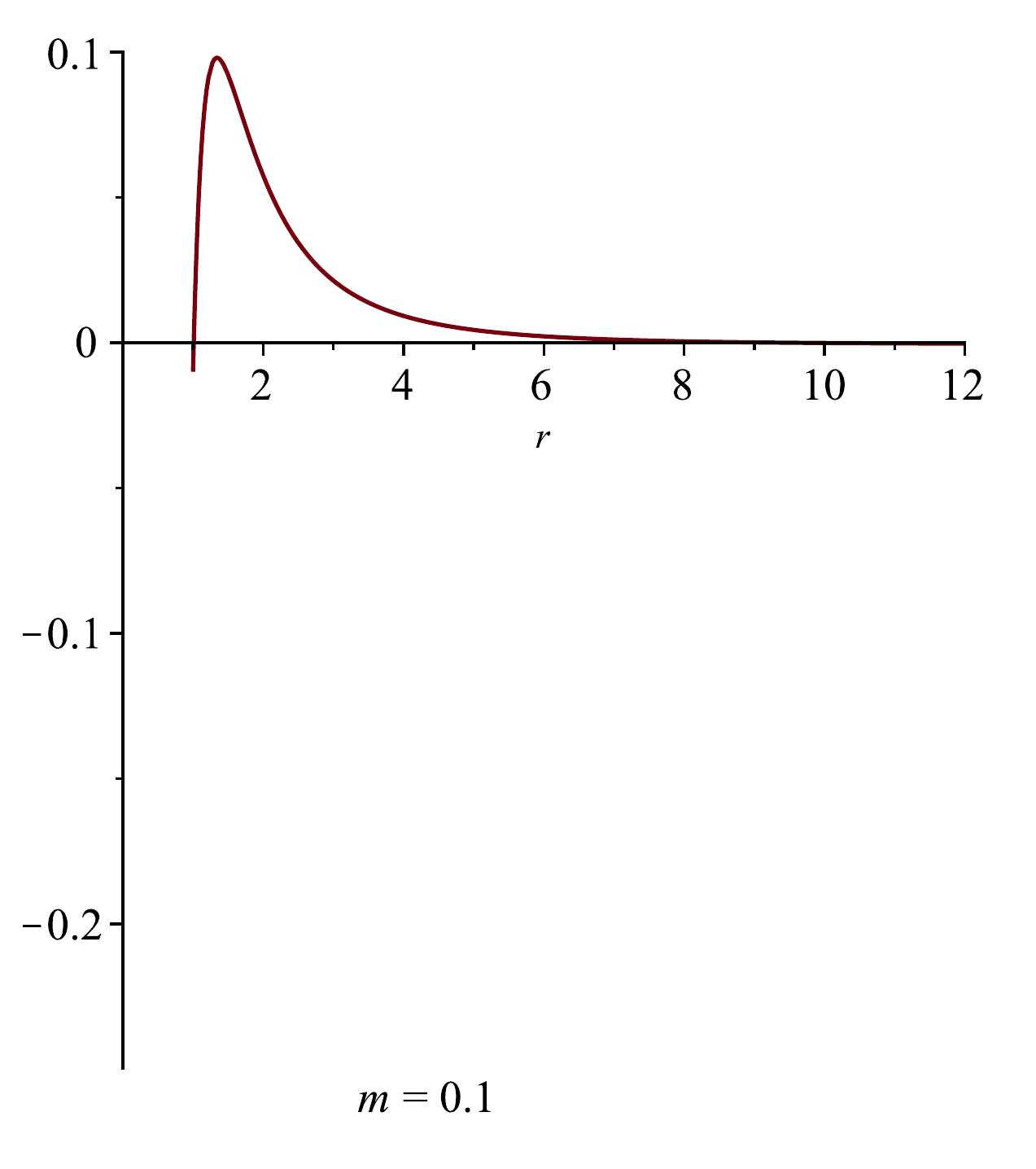} \includegraphics[width=6cm]{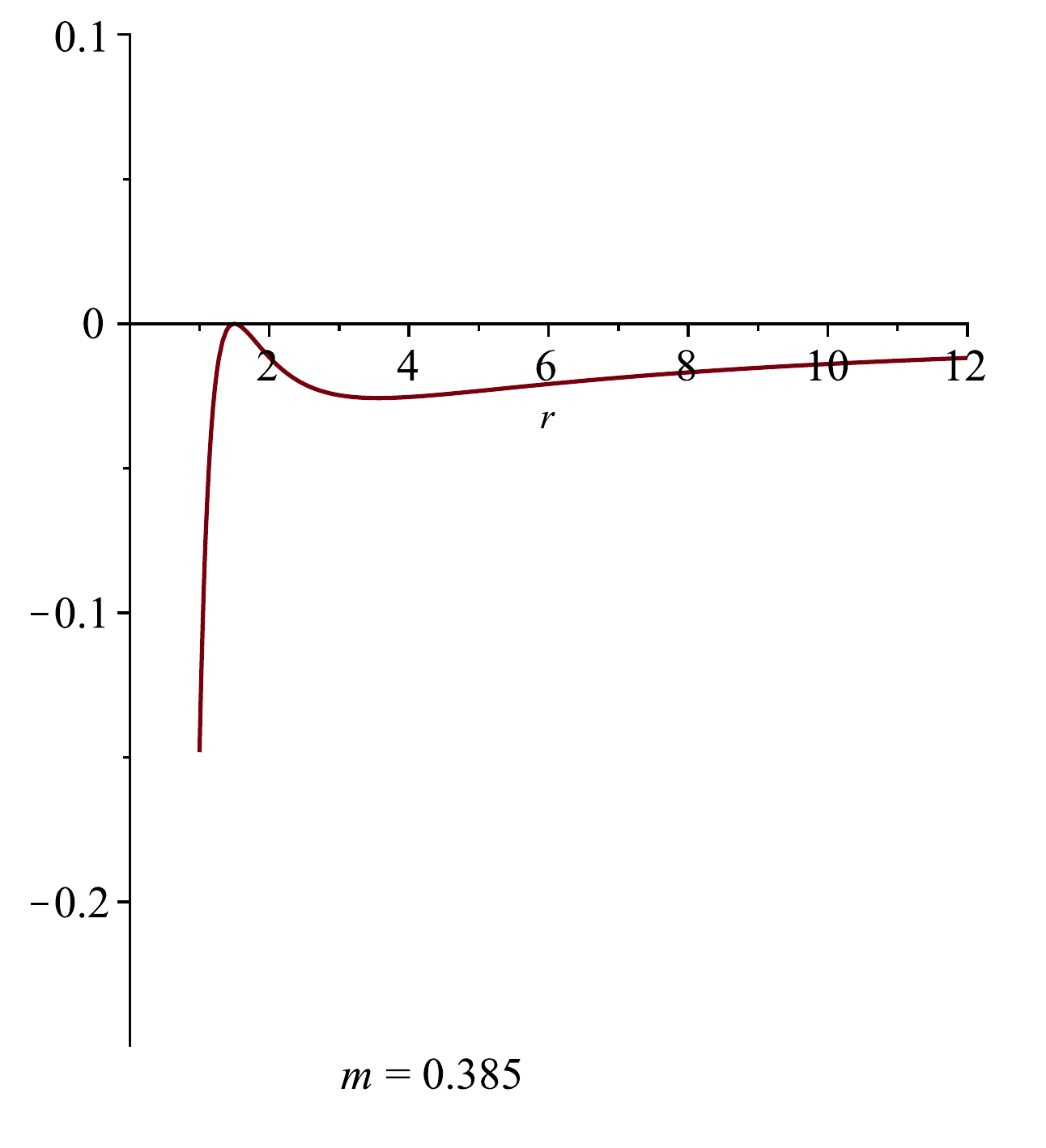} \includegraphics[width=6cm]{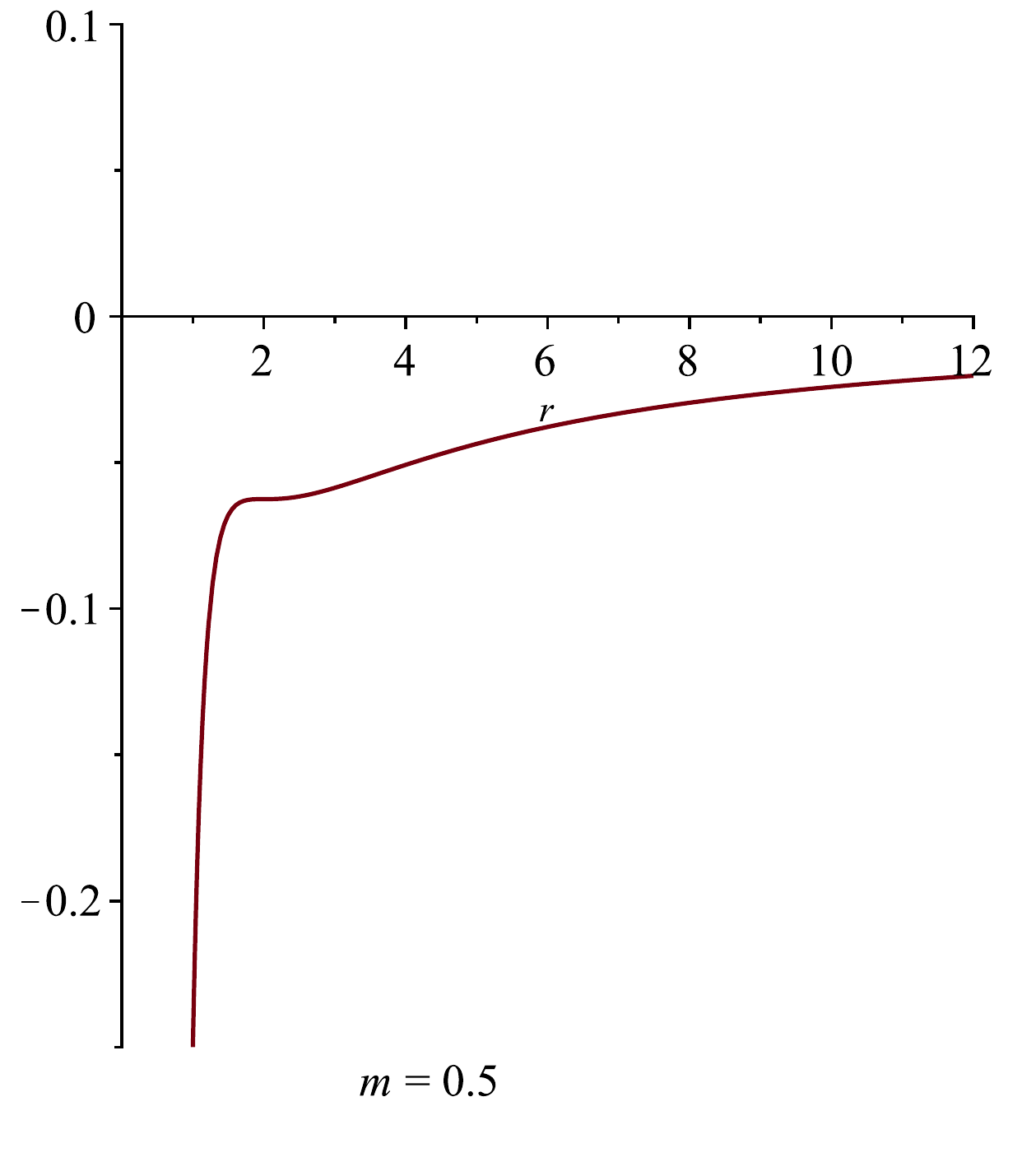}}
\caption{The effective potential $- {m^2 \over r} + {1 \over r^3} - {1 \over r^4}$ in the region outside the horizon $r>1$, plotted for three different values of $m$.  From left to right $m = 0.1,\,0.385,\,0.5$.\label{fig:potential}}
\end{figure}

Although (\ref{generalsolution}) is the general solution, it is not a particularly transparent formula.  So in what follows we will develop approximations which are valid in different mass ranges.
To gain some intuition before proceeding it is useful to return to the differential equation (\ref{KGeq2}) and recast it in the form
\begin{equation}
\label{Schrodinger}
\left(-\partial_{r_*}^2 - {m^2 \over r} + {1 \over r^3} - {1 \over r^4} \right) R = \bar{k}^2 R \, .
\end{equation}
Here we are setting $\phi(t,r) = e^{-i \omega t} {1 \over r} R(r)$.  This resembles a Schr\"odinger equation in a potential $- {m^2 \over r} + {1 \over r^3} - {1 \over r^4}$
that is shown Fig. \ref{fig:potential}.  If $m$ is small there is an ${\cal O}(1)$ potential barrier around the unstable maximum at $r \approx {4 \over 3}$.
As $m$ increases the barrier comes down.  When $m = {2 \over 3 \sqrt{3}} \approx 0.385$ the top of the barrier is at zero, and for large $m$ the potential is
purely attractive.  The physics is then pretty clear.  At large radius a wave could be coming in from the right.  For small $m$ and sufficiently low energy the incoming wave will reflect off the barrier, with a small
probability of tunneling into the black hole.  But if $m$ (or the energy) is large enough the incoming wave will be almost completely absorbed by the black hole.  The black hole itself corresponds to a boundary condition,
namely that at the horizon $r \rightarrow 1$ the wave must be purely ingoing (into the horizon).

Now let us return to the general solution (\ref{generalsolution}).
The definition and properties of the confluent Heun function ${\sf
  HeunC}$ are given in appendix \ref{appendix:asymptotics}.
In the near-horizon limit $r \rightarrow 1$, {\sf HeunC} approaches
unity. Thus, the constant $c_1$ should be set to zero, so that $\phi$
is purely ingoing at the horizon i.e.
\begin{equation}
\label{NearHorizon0}
\phi(t,r) \overset{r \to 1 }{\approx}  c_2 \, e^{-i \omega (t + r_*)} e^{i(\omega - \bar{k})} \, .
\end{equation}
In other words, from now on, we consider:
\begin{equation}
\label{smooth}
\phi(t,r) = 
c_2 \, e^{-i \omega t} (r - 1)^{-i \omega} e^{-i \bar{k} r} \, {\sf HeunC}(2i\bar{k}, -2i\omega, 0, -\omega^2 -\bar{k}^2, \omega^2 + \bar{k}^2, 1 - r) \, ,
\end{equation}
where $c_2$ is an integration constant whose size will be determined
by the dark matter density far away from the black hole.

We are interested in particles that are gravitationally bound within the galaxy but have a small binding energy $E < 0$.  That is, we are interested in $\omega$ slightly smaller than $m$.  For simplicity to capture the relevant physics we will focus on the marginally-bound case $\omega = m$.  From (\ref{Schrodinger}) this should be a reasonable approximation as long as $\bar{k}^2$ is small compared to the potential, $\vert \bar{k} \vert^2 < m^2 / r_i$.  So we set $\omega = m$ and $\bar{k} = 0$ and take
\begin{equation}
\label{k=0}
\phi(t,r) = c_2 \,  e^{-i m( t+\log(r - 1))} \, {\sf HeunC}(0,-2im,0,-m^2,m^2,1-r) \, .
\end{equation}

Note that for the sake of generality we are allowing $\phi$ to be complex.
However for many applications $\phi$ is real (e.g.\ if $\phi$ is the angular field
corresponding to an axionic degree of freedom).  To treat the two cases in parallel
we adopt the convention that a real solution can be obtained from a complex solution
by adding the complex conjugate.  So for example we take the real field corresponding to (\ref{k=0})
to be
\begin{equation}
\label{realphigeneral}
\phi(t,r) =
c_2 \,  e^{-i m( t+\log(r - 1))} \, {\sf HeunC}(0,-2im,0,-m^2,m^2,1-r) 
+ \, {\rm c. \, c.\,} \, ,
\end{equation}
where ${\rm c. c.}$ stands for complex conjugation.  This convention fixes the relative normalization of what we mean by $c_2$ in the real and complex cases.
It is a useful convention for reasons we turn to next.    {\it For the most part, the fields below are given for a
complex field $\phi$, but it is simply a matter of adding the complex conjugate if one is interested in a real scalar $\phi$.}  

Before proceeding let us discuss our procedure for normalizing the field amplitude.  It is convenient to fix the normalization in terms of a physical observable, namely the energy density
of the scalar field $\rho_i$ evaluated at a radius $r_i$ which is much larger than the Schwarzschild radius.  We have in mind that $r_i$ is the radius of the sphere of impact; see the discussion around (\ref{ridef})
for a precise definition of this quantity.  First consider a complex scalar field.  To evaluate the energy density we use the stress tensor given in appendix \ref{app:flux}.  Far from the black hole (so that the geometry
is approximately flat) and neglecting spatial gradients (as appropriate for non-relativistic particles) the energy density of a complex scalar field is
\begin{equation}
\rho = T^{tt} \approx \vert \partial_t \phi \vert^2 + m^2 \vert \phi \vert^2 \approx 2 m^2 \vert \phi \vert^2 ,
\end{equation}
where we assumed $\omega \approx m$.  Thus for a complex scalar field we fix the field amplitude by setting
\begin{equation}
\rho_i = 2 m^2 \vert \phi(r = r_i) \vert^2 .
\end{equation}
Now let us consider a real scalar field.  Given a complex scalar $\phi = A e^{-i m t}$ where $A$ is a real amplitude, we would build a real scalar field
by setting $\phi = A e^{-i m t} + {\,\rm c. c.} = 2 A \cos m t$.  For a real scalar the energy density has a factor of $1/2$,
\begin{equation}
\label{rho}
\rho \approx {1 \over 2} \big(\dot{\phi}^2 + m^2 \phi^2\big) = 2 m^2 A^2 .
\end{equation}
This means the real scalar field we build by adding the complex conjugate has exactly the same energy density as the complex scalar we started from.  {\it Below we will write formulas for complex
fields, normalized by their energy density $\rho_i$.  To obtain a real field merely add the complex conjugate to the expressions below; $\rho_i$ will be the energy density of the real scalar field at $r_i$.}

So far we have found the exact solution (\ref{k=0}), which near the horizon becomes an ingoing wave
\begin{equation}
\label{NearHorizonk=0}
\phi(t,r) = c_2 \,  e^{-i m( t+\log(r - 1))} \quad \hbox{\rm as $r \rightarrow 1$} .
\end{equation}
Now let us consider the large $r$ behavior of $\phi(t,r)$. 
The large $r$ limit of the confluent Heun function (with $\bar k =
0$) is derived in appendix \ref{app:KG} and implies\footnote{Note that the large $r$ behavior for a non-zero $\bar k$ is
  quite different. See Eq. \eqref{asymphiso}.}
\begin{equation}
\label{LargeRadius}
\begin{split}
\phi (t,r) & =  c_3 \,  e^{-i m t} \, {1 \over r^{3/4}} e^{2 i m \sqrt{r}} \left(1 + {\cal O}\left(\frac{1}{m\sqrt{r}}\right)\right) \\
& \quad +  c_4 \,  e^{-i m t} \, {1 \over r^{3/4}} e^{-2 i m \sqrt{r}} \left(1 + {\cal O}\left(\frac{1}{m\sqrt{r}}\right)\right) \, ,
\end{split}
\end{equation}
where $c_3$ and $c_4$ are constants related to $c_2$ (the precise
relations to be given below).  This has a simple interpretation, that $c_3$ is the coefficient of an outgoing wave
and $c_4$ is the coefficient of an ingoing wave.
This is a good approximation as long as $m \sqrt{r} \gg 1$.
In other words, this is not simply a large $r$ expansion, but rather
a large $m \sqrt{r}$ expansion. Restoring $r_s$, the expansion is valid for $m \gg 1/\sqrt{r r_s}$. 
The reader might wonder where the $r^{-3/4}$ behavior comes from, and why there is a $\sqrt{r}$ in the exponent:
an intuitive explanation is given in Sec. \ref{sec:conclude}.
An expression like Eq. (\ref{LargeRadius}), while strictly valid only
for $m \gg 1/\sqrt{r}$, will be treated as an acceptable approximation
for $m \, \gsim \, 1/\sqrt{r}$ with the understanding that when $m$
approaches $1/\sqrt{r}$, there would be order one corrections.

We now summarize the behavior of the field in different regimes.  To start suppose the mass is such that the combination
$m \sqrt{r_i}$ is large, where $r_i$ is the radius of the sphere of impact.  In this case the field near $r_i$ is described by (\ref{LargeRadius}).
This large $m\sqrt{r_i}$ limit can be further divided into two separate regimes.
One is where the mass is so large that $m \, \gsim \, 1$.  In this case, which we call regime IV, there is no potential barrier around the black hole
as can be seen in the right panel of Fig.\ \ref{fig:potential}.  Thus we expect to have an ingoing wave everywhere.
Indeed the relevant approximation for the Heun function is given in (\ref{HCbig}).  It implies that the large-$r$ behavior is given by
setting $c_3 \approx 0$ and $c_4 \approx c_2$ in (\ref{LargeRadius}), and that (\ref{LargeRadius}) is in fact a good approximation
as long as $r \gtrsim r_s$.  Very close to the horizon we still have the near-horizon behavior (\ref{NearHorizonk=0}).  That is,
\vspace{.15cm}
\begin{tcolorbox}[colframe=white,arc=0pt,colback=greyish2]
\vspace{-.42cm}
\begin{eqnarray}
\label{equation:RegimeIV}
&& {\rm Regime \, \, IV:} \,\, m \,\, \gsim \,\, r_s^{-1} \nonumber \\
&& \phi (t, r) \approx \sqrt{\rho_i \over 2 m^2} \left( {r_i \over r}
  \right)^{3/4} e^{-imt} e^{-i 2 m\sqrt{r r_s}} \quad {\rm
  for\quad} r_s \, \lsim \,  r < r_i \, , \nonumber \\
     \label{RegimeIVsol}
&& \phi (t, r) \approx c_2 \, e^{-i m (t + r_*)} e^{im r_s}
\quad {\rm where\,\,} c_2
   \approx \sqrt{\rho_i \over 2 m^2} \left({r_i \over r_s}\right)^{3/4}
\quad {\rm for\quad} r \rightarrow r_s \, ,
\end{eqnarray}
\end{tcolorbox}
where $r_s$ is restored for clarity.
We fixed the normalization of $\phi$ by setting the energy density to be $\rho_i$
at $r = r_i$.  The large $r$ expression is strictly correct only for $r \gg r_s$ but
is in practice a reasonable approximation down to $r \gtrsim r_s$.
Thus the amplitude at the horizon $c_2$ is given by extrapolating the large $r$ expression to $r = r_s$.

Next we consider the case where $m \, \lsim \, 1$, while $m \, \gsim \, 1/\sqrt{r_i}$ continues to hold.  In this case, which we refer to as regime III,
the field near $r_i$ is well-described by (\ref{LargeRadius}).  Thus we have ingoing and outgoing waves near the sphere of impact.  However since $m \lsim 1$
there is a significant potential barrier near the black hole, as seen in the left panel of Fig.\ \ref{fig:potential}, and we expect the ingoing wave to reflect off the barrier.  That is, we expect $c_3$ and $c_4$ to have
roughly the same magnitude.  This means a standing wave will develop around the black hole.  The relevant approximation is given in (\ref{RIII}) and implies, restoring $r_s$ for clarity:
\vspace{.15cm}
\begin{tcolorbox}[colframe=white,arc=0pt,colback=greyish2]
\vspace{-.42cm}
\begin{eqnarray}
&& {\rm Regime \, \, III:} \,\, (r_i r_s)^{-1/2} \,\,\lsim\,\, m \,\, \lsim \,\, r_s^{-1} \nonumber \\
&& \phi (t, r) \approx \sqrt{\rho_i \over m^2} \left( {r_i \over r}
  \right)^{3/4} e^{-imt} {\,\rm cos\,} \left(2m \sqrt{r r_s} -
   3\pi/4\right)
\quad {\rm for} \quad    m^{-2} r_s^{-1} \, \lsim \, r < r_i \, ,
   \nonumber \\
   \label{RegimeIIIsol}
&& \phi (t, r) \approx c_2 \, e^{-i m (t + r_* - r)} \quad {\rm where\,\,}
c_2 \approx \sqrt{\rho_i \over m^2} \left({r_i \over m^{-2} r_s^{-1}}\right)^{3/4} \sqrt{\pi} \quad {\rm for\quad} r_s < r \, \lsim \, m^{-2} r_s^{-1}. \nonumber \\
\end{eqnarray}
\end{tcolorbox}
Note that the energy density $\rho$ oscillates in space like 
$[{\rm cos\,} (2 m \sqrt{r r_s} - 3\pi/4)]^2$; the quantity $\rho_i$ is not necessarily the
density at precisely $r_i$, but is rather the density averaged over
an oscillation cycle around $r_i$ i.e.\ replacing the $[{\,\rm cos\,}]^2$ by
$1/2$.  As explained in the appendix, this result is obtained assuming $m \ll 1$, but it gives a reasonable
approximation as long as $m \lesssim 1$.  A noteworthy point about the profile in regime III:
aside from the oscillations which pile up against the horizon, the field is roughly constant from $r_s$ out to $1/m^2r_s$, then begins to drop off
like $1/r^{3/4}$.

Now let us consider what happens when $m \, \lsim \, 1/\sqrt{r_i}$. As shown in Appendix
\ref{appendix:asymptotics}, the Heun function has an expansion in the
small mass limit:\footnote{Restoring units, this expansion in powers
of $m$ is valid if both $m r_s \ll 1$ and -- more importantly -- if $m \sqrt{r r_s} \ll 1$.}
\begin{equation}
\label{expansion}
{\sf HeunC}(0,-2i m,0,-m^2, m^2, z ) = 1 + i m {\,\rm log\,} (1 - z) + {\cal O}(m^2) \, .
\end{equation}
Therefore, in this limit, $\phi(t,r) $ takes the form
(see Appendix \ref{appendix:asymptotics}):
\begin{eqnarray}
\label{smallmExpand}
\phi(t, r) = c_2 \,  e^{ -i m t } \left(  1-i m {\,\rm log\,} \left( 1- r^{-1}
  \right) - {1\over 2} m^2 r + ...  \right)  \, .
\end{eqnarray}
A few comments are in order about this expression. 
First of all, we are not displaying all the order $m^2$ terms in the
parentheses; they can be found in Appendix
\ref{appendix:asymptotics}.  We keep the $m^2 r$ term because among
the $m^2$ terms, this dominates at large $r$. There is a delicate
balance here, we are interested in large $r$, but not so large that
$m^2 r \, \gsim \, 1$,  i.e. we are interested in $m^2 r \, \lsim \, 1$, or
$m \,\lsim \,1/\sqrt{r}$, or $r \, \lsim \, 1/m^2$.\footnote{Note that as long as $m \, \lsim \, 1/\sqrt{r}$, 
$m \, \lsim \, 1$ is automatic because $r \, \gsim \, 1$, implying
the logarithm term is also small compared to unity.
Moreover, $m \,\lsim \, 1/\sqrt{r}$ also implies $m \,\lsim\, r$,
consistent again with the fact that the logarithm term is small
compared to unity (recall that the logarithm can be expanded to give
$i m / r$ for large $r$). In other words Eq. (\ref{smallmExpand}) can be thought of
as a small $m \sqrt{r}$ but large $r$ expansion.}
Now, for large $r$, the logarithm can
be expanded to give a term that goes like $i m /r$. 
Comparing these two terms, we see that whether one dominates over the other
is determined by how large $m$ is compared to $1/r^2$, or how big $r$
is compared $1/\sqrt{m}$. 
There are thus roughly two possibilities.
Suppose we are in what we will call regime II: 
$1/r_i^2 \, \lsim \, m \, \lsim \, 1/\sqrt{r_i}$---recall that
$r_i$ is the largest radius we can go out to before the geometry
deviates from Schwarzschild---then, for $1/\sqrt{m} \,\lsim\, r
\,\lsim\, r_i$, the field profile is dominated by the $m^2r$ term\footnote{Even in this case, it is useful to keep the logarithm or
  $im/r$ term because it contributes to the energy flux in a
  non-trivial way (see below).}, but
for $1 \,\lsim \, r \,\lsim \, 1/\sqrt{m}$, the field profile is
dominated by the logarithm or $im/r$ term.
The second possibility is what we will call regime I:
$m \,\lsim \, 1/r_i^2$. In this case, the logarithm or $im/r$ term always dominates
over the $m^2 r$ term (because $r \,\lsim \, r_i$ to stay within the
Schwarzschild geometry). Summarizing, we have, again restoring $r_s$:
\vspace{.15cm}
\begin{tcolorbox}[colframe=white,arc=0pt,colback=greyish2]
\vspace{-.42cm}
\begin{eqnarray}
&& {\rm Regime \,\, II:} \,\, r_s / r_i^2 \,\, \lsim\,\,  m \,\, \lsim
   \, \, 1/\sqrt{r_i r_s} \nonumber \\
&& \phi (t, r) \approx \sqrt{\rho_i \over 2 m^2} e^{-imt} \left(1 + {im r_s^2
   \over r} - {1\over 2} m^2 r r_s \right) \quad {\rm for \quad}
   \sqrt{r_s/m} \,\, \lsim \,\, r
   \,\,\lsim r_i \, , \nonumber \\
&& \phi (t, r) \approx \sqrt{\rho_i \over 2 m^2} e^{-imt} \left(1 + {im r_s^2
   \over r} \right) \quad {\rm for \quad}
   r_s \,\, \ll \,\, r \,\, \lsim \,\, \sqrt{r_s/m} \nonumber \\
   \label{smallm1}
&& \phi (t, r) \approx c_2 \, e^{-i m (t + r_*)} \quad {\rm where\,\,}
c_2 \approx \sqrt{\rho_i \over 2 m^2} \quad {\rm for\quad} r \rightarrow
   r_s \, ,
\end{eqnarray}
\end{tcolorbox}
and
\vspace{.15cm}
\begin{tcolorbox}[colframe=white,arc=0pt,colback=greyish2]
\vspace{-.42cm}
\begin{eqnarray}
&& {\rm Regime \,\, I:}  \,\, m \,\, \lsim \,\, r_s / r_i^2 \nonumber \\
&& \phi (t, r) \approx \sqrt{\rho_i \over 2 m^2} e^{-imt} \left(1 + {im r_s^2
   \over r} \right) \quad {\rm for \quad}
   r_s \,\, \ll \,\, r \,\, \lsim \,\, r_i \nonumber \\
   \label{smallm2}
&& \phi (t, r) \approx c_2 \, e^{-i m (t + r_*)} \quad {\rm where\,\,}
c_2 \approx \sqrt{\rho_i \over 2 m^2} \quad {\rm for\quad} r \rightarrow
   r_s \, .
\end{eqnarray}
\end{tcolorbox}
In regime I, which is the extreme small mass limit (or the wave limit),
one can make a stronger statement about the
scalar profile. Ignoring the order $m^2$ term in
both the spatial and temporal dependence in
Eq. (\ref{smallmExpand}), but without expanding the logarithm, we have
(restoring $r_s$):
\begin{equation}
\label{JacobsonHair}
\phi(t, r) = c_2 \,  \left(  1-i m \left[t +  r_s {\,\rm log\,} \left( 1-
      {r_s \over r} \right) \right] \right)  \, .
\end{equation}
The zero order solution $\phi(t, r) = c_2$ is of course the trivial
solution to the massless Klein-Gordon equation. The order $m$ solution
is in fact the non-trivial solution to the massless Klein-Gordon
equation found by Jacobson:
$\phi \propto (t + r_s {\,\rm log\,} (1 - r_s/r))$
(Eq. (\ref{jacsol}))). It is worth emphasizing that
this (massless) solution holds at all radii.

It is useful to give an example of what a real scalar field profile
looks like.
For instance, in regime I, by adding the complex conjugate, one obtains
in the radius range $r_s \ll r \,\,\lsim\,\,
r_i$:
\begin{eqnarray}
\phi (t, r) \approx \sqrt{2 \rho_i \over m^2} 
\left( {\,\rm cos\,} (mt)  + {m r_s^2 \over r} {\,\rm sin\,} (mt)
  \right) \, .
\label{Unruh}
\end{eqnarray}
Note that the time origin is arbitrary; for instance, one could swap
the cosine and the sine by shifting $t$. Note also that by assumption,
regime I implies $m r_s \ll 1$ (see Fig.\ \ref{Fig:1}) and so the
$1/r$ scalar hair is weak. In other words, the scalar field amplitude
at the horizon is not too different from the amplitude far away (at
$r_i$). The $1/r$ tail is important for getting the correct energy
flux, however, as we will see.\footnote{It is also worth emphasizing that this $1/r$ tail was at some level
known in much earlier work, see e.g. equation 42 of \cite{Unruh:1976fm}
(also \cite{Detweiler:1980uk}). What we have done here is rather
modest---an exploration of the scalar profile or hair as the mass is varied,
using properties of the Heun function. The use of the Heun function
and its expansion (\ref{smallmExpand}) also allows us to deduce the $1/r$ tail
with the correct coefficient without going through matching
procedures.}

\begin{figure}[ht]
\begin{center}
\includegraphics[width=12cm]{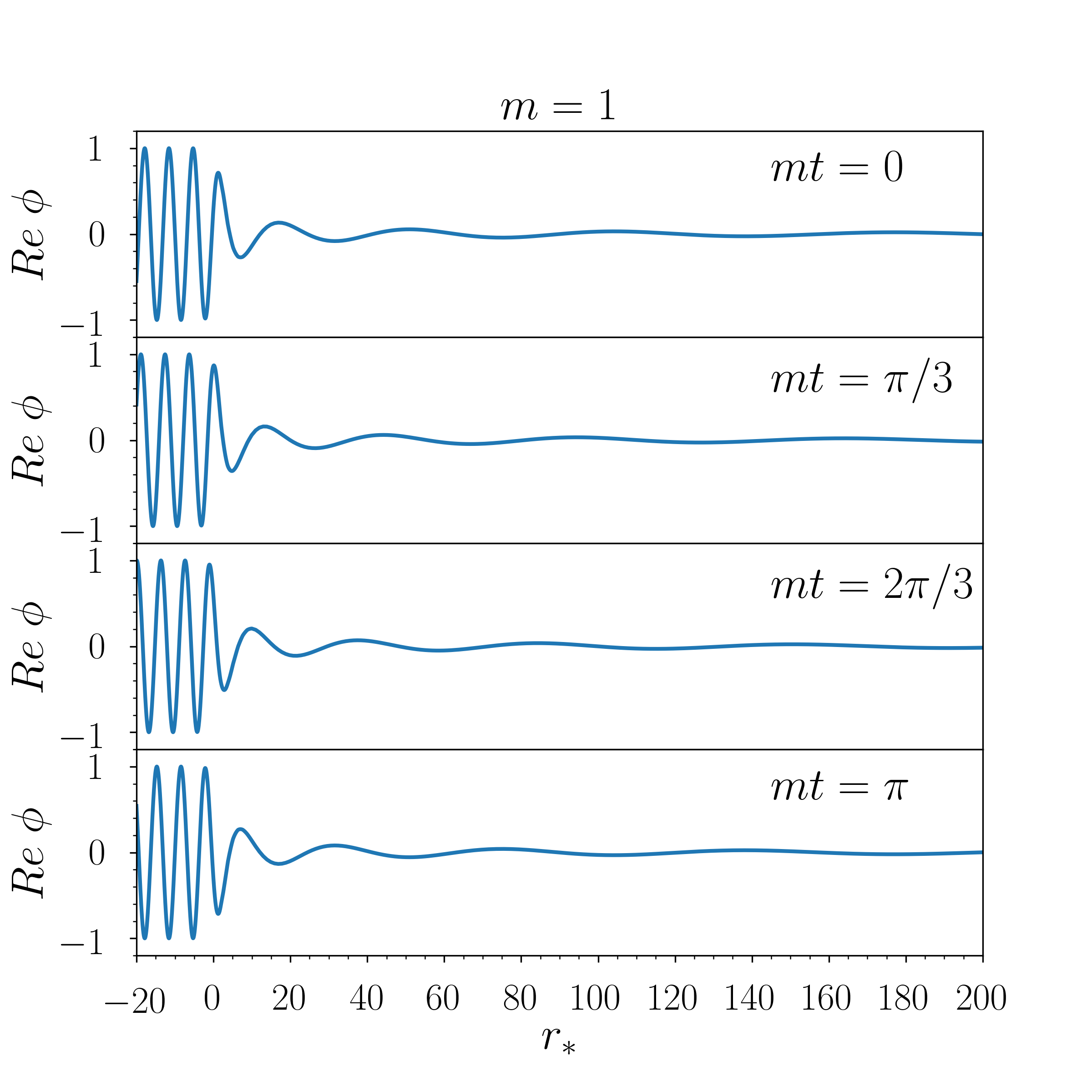}
\end{center}
\caption{Four snapshots of a real scalar field with $m = 1$ as a function of $r_*$.  These are frames from an animation.  To view the video
on arXiv.org follow the link under ``Ancillary files''.  Here $\phi$ is normalized to unity at the horizon.
\label{fig:movie}}
\end{figure}

\begin{figure}
 \centering
      \includegraphics[width=0.49\columnwidth]{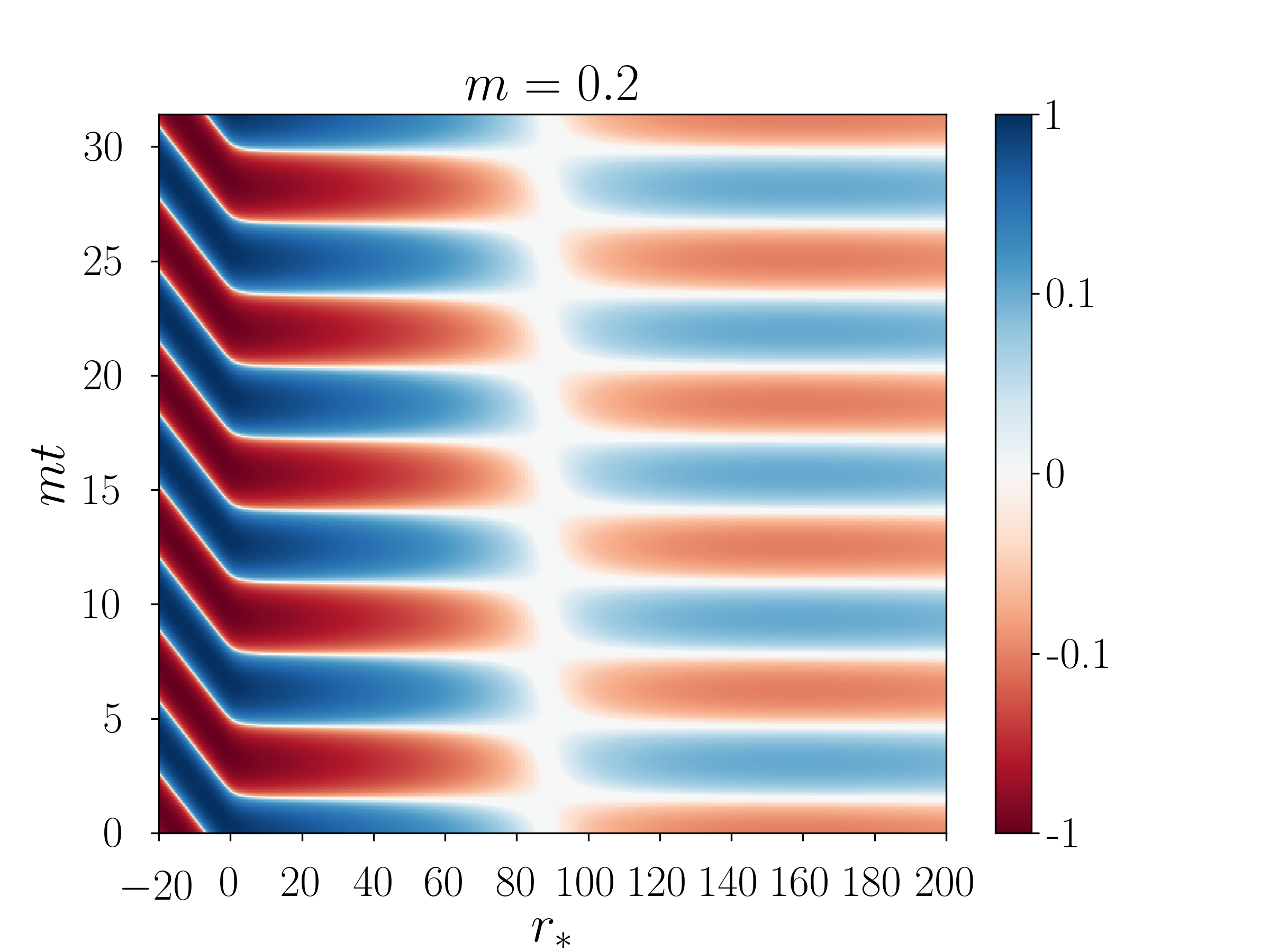}
        \includegraphics[width=0.49\columnwidth]{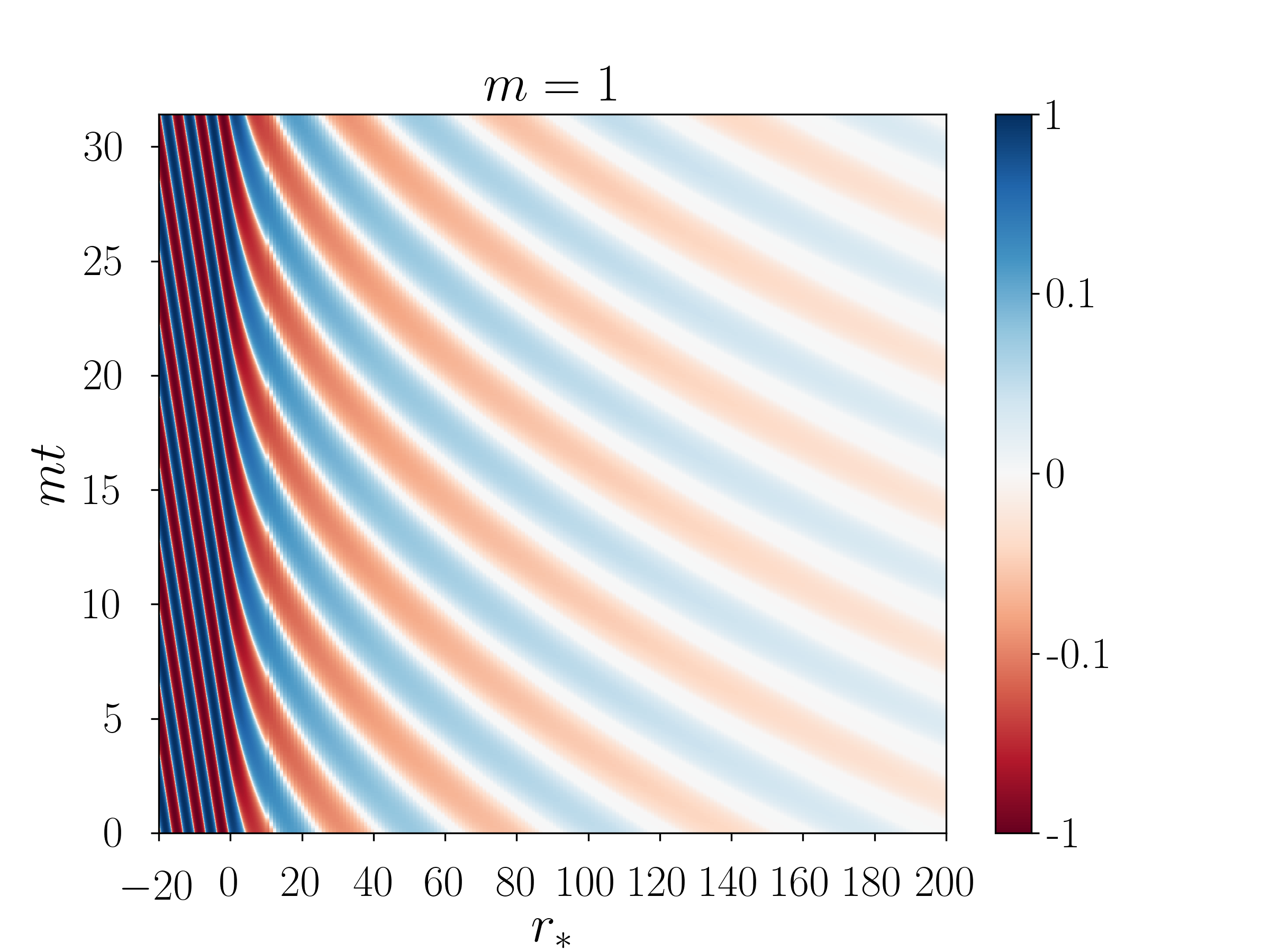}
\caption{Two massive real scalar fields (left: $m=\frac{1}{5}$; right:
  $m=1$ ) in a pure Schwarzschild geometry ($r_s=1$), showing $\phi$
  as a function of $t$ and $r_{*}$ where $r_*$ is the tortoise
  coordinate. The color scheme shows the scalar field amplitude
  with the largest positive value as deep blue and the most negative
  value as deep red. One can see that the scalar field oscillates with
  a larger amplitude closer to the horizon.
  At $r_* >0$, a heavy scalar (regime IV) is purely
  ingoing due to the vanishing barrier in the potential.
  At $r_* > 0$, the lighter scalar (regime III) has both ingoing and
  outgoing waves of comparable amplitudes, hence the standing wave
  pattern with a node at $r_* \sim 90$.
For $r_{*} <0$, the potential barrier almost vanishes therefore $\phi$
becomes purely ingoing plane wave with the speed of light.
(We have effectively chosen $r_i \sim 200$ in these illustrations.)
 \label{fig:massive_phi}}
\end{figure}

\begin{figure}
\centerline{\includegraphics[width=0.7\columnwidth]{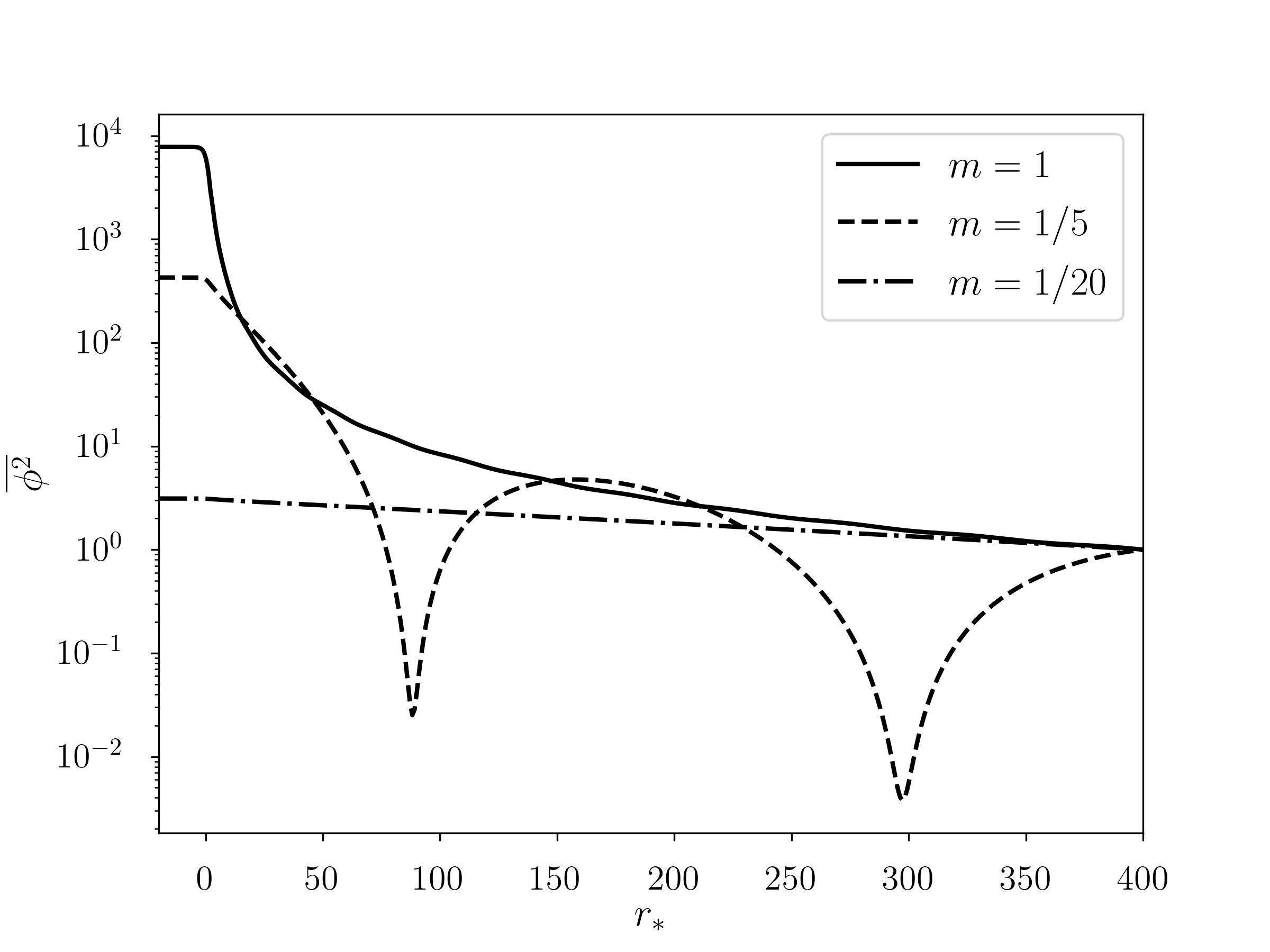} }
\caption{This shows the time-averaged $\overbar{\phi^2}$ as a function of the tortoise
  coordinate $r_*$ in a Schwarzschild geometry with $r_s = 1$. 
The curves are all normalized to unity at $r_* = 400$
i.e.\ $r_i$ is effectively chosen to be $\sim 400$.
The masses of $1$, $1/5$ and $1/20$ roughly span
regimes IV, III and II.}
\label{fig:avphi}
\end{figure}

Let us illustrate the different behaviors of the Klein Gordon
scalar by some numerical examples.  Fig.\ \ref{fig:movie} shows the
field as a function of $t$ and $r_*$ for $m = 1$ (roughly regime IV).  It illustrates the
growth of the field near the black hole and the plane wave propagating into the horizon.
Fig. \ref{fig:massive_phi} shows the real scalar $\phi$ as a function of
$t$ and $r_*$, for two different masses (in essentially regimes III and
regime IV). In Figs.\ \ref{fig:movie} and \ref{fig:massive_phi} (and
only in those figures) we fixed the field amplitude to be 1 at the
horizon, 
that is, we set $c_2 = 1$.
Finally, to see how the amplitude of the field varies with radius for different masses, it is convenient to eliminate
the time dependence by time-averaging $\phi^2$ over a period of oscillation.  The resulting $\overbar{\phi^2}$
is shown in Fig. \ref{fig:avphi}.  The masses of $1$, $1/5$ and $1/20$ roughly span
regimes IV, III and II, and one can see pile-up close to the horizon
in IV and III, the standing wave in III and a rather flat profile for
II, in agreement with the analytic approximations given above.  We
should emphasize that all these figures (Figs. \ref{fig:movie},
\ref{fig:massive_phi} 
and \ref{fig:avphi}) show numerical solutions
to the differential equation, obtained using the {\sf HeunC} function in Maple, and do not rely on any approximations.

Lastly, we conclude this section by computing the energy flux near the
horizon.
Conservation of energy-momentum can be expressed as
\begin{eqnarray}
{1\over \sqrt{-g}} \partial_\mu (\sqrt{-g} T^\mu {}_\nu) - {1\over 2}
( \partial_\nu g_{\mu\lambda} ) \, T^{\mu\lambda} = 0\, ,
\end{eqnarray}
which uses the fact that the energy-momentum tensor $T^{\mu\lambda}$ is symmetric.
 Because the Schwarzschild metric is time
independent, the conservation of energy (for $\phi$ with only $t$ and
$r$ dependence) takes a particularly simple
form:
\begin{eqnarray}
\label{Tconserve}
\partial_t (r^2 T^t {}_t) + \partial_r (r^2 T^r {}_t) = 0 \, .
\end{eqnarray}
The amount of energy flowing inward across a sphere of radius $r$ per
unit time is thus
\begin{eqnarray}
\Phi = 4\pi r^2 T^r {}_t = - 4 \pi r^2 \left(1-\frac{1}{r} \right) T^{rt}\, .
\end{eqnarray}
The energy-momentum tensor is given in \eqref{tmunu} for a complex scalar and (\ref{Treal}) for a real scalar.
So more explicitly we have
\begin{eqnarray}
\Phi = 4\pi r^2 \left(1 - {1\over r}\right) \left[ \partial_t
  \phi^* \partial_r \phi + \partial_t \phi \partial_r \phi^* \right]
  \, 
\end{eqnarray}
for a complex scalar and
\begin{eqnarray}
\Phi = 4\pi r^2 \left(1 - {1\over r}\right) \partial_t
  \phi \partial_r \phi  \, 
\end{eqnarray}
for a real scalar.
For a complex scalar with the $e^{-i\omega t}$ time dependence, it is
manifest that both $T^t {}_t$ and $T^r {}_t$ would be
time-independent. Thus, Eq. (\ref{Tconserve}) tells us that $\Phi$
should be independent of $r$. For a real scalar, which would have 
${\,\rm cos} (\omega t)$ or ${\,\rm\, sin} (\omega t)$ time
dependence, $T^t {}_t$ and $T^r {}_t$ are no longer time independent.
However, as long as one averages over a cycle of oscillation, 
it can be shown that $\partial_t T^t {}_t = 0$, in which case
Eq. (\ref{Tconserve}) also tells us the averaged $\Phi$ is $r$
independent.

In other words, $\Phi$ (or its average) can be computed at any radius
$r$ including in particular the horizon. Using expressions in
appendix \ref{app:flux}, and the near-horizon solution 
Eq. (\ref{NearHorizon0}), we find
\begin{equation}
\label{HorizonFlux0}
\Phi = 8 \pi \omega^2 \vert c_2 \, e^{-i \bar{k}}\vert^2 \, ,
\end{equation}
where we keep $\omega$ and $\bar k$ general, even though we will mostly be interested in $\omega = m$ and $\bar{k} = 0$.
For a real scalar $\phi$, with the convention described below (\ref{realphigeneral}), we have
\begin{eqnarray}
\Phi_{{\rm real }\; \phi} = 16 \pi \omega^2 | c_2 \, e^{-i \bar{k}} |^2 \sin^2 \omega t  \, .
\end{eqnarray}
Note that $c_2 e^{-i\bar k}$ in principle has a phase that can enter
into the phase of the sine which we ignore because the origin of $t$
is arbitrary in any case. Averaging over time: 
\begin{eqnarray}
{\rm time \,\, averaged\,\,} \Phi_{{\rm real}\; \phi} = 8 \pi \omega^2 | c_2 \,
  e^{-i \bar{k}} |^2 \, .
\end{eqnarray}
At the horizon, it does not seem so crucial that we average over
time. But it can be checked easily that at larger radii, it is only the
time averaged $\Phi$ that is $r$ independent for a real scalar
(try for instance computing $\Phi$ for Eq. (\ref{Unruh})). 

Given our results for the scalar field it is straightforward to relate the energy accretion rate $\Phi$ to the energy density at the sphere of impact $\rho_i$.
Setting $\bar{k} = 0$, so that $\omega = m$ and $\Phi = 8 \pi m^2 \vert c_2 \vert^2$, we find that, as follows from \cite{Unruh:1976fm},
\begin{eqnarray}
\label{PhiOneTwo}
&& \hbox{\rm regimes I and II ($m \lesssim 1/\sqrt{r_ir_s}$):} \quad \Phi = 4 \pi r_s^2 \rho_i  \, ,\\
&& \hbox{\rm regime III ($1/\sqrt{r_ir_s} \lesssim m \lesssim 1/r_s$):} \quad \Phi = 2\pi \cdot 4\pi r_s^2 \rho_i \left(r_i \over m^{-2} r_s^{-1}\right)^{3/2} \, , \\
\label{PhiFour}
&& \hbox{\rm regime IV ($m \gtrsim 1/r_s$):} \quad \Phi = 4 \pi r_s^2 \rho_i \left(r_i \over r_s\right)^{3/2} \, .
\end{eqnarray}
A few comments on these results are in order.  First, in regimes I and II where the mass is small, the field amplitude and hence the energy density is roughly constant from $r_i$ down to the
horizon.\footnote{This is discussed in more detail in section \ref{sect:accretion}.}  Near the horizon the field is effectively massless and waves move toward the horizon at the speed of light.  To emphasize this point we can restore the speed of light and write the
flux in regimes I and II as $\Phi = 4 \pi r_s^2 \rho_i c$.  In regime III there is an enhancement factor associated with the $1/r^{3/4}$ growth of the field from $r_i$ down to the radius $1/m^2r_s$
where the growth cuts off.  The numerical coefficient in regime III is a bit ambiguous since the energy density oscillates in space; we have adopted the averaging procedure described below (\ref{RegimeIIIsol}) to get a rough estimate.
In regime IV there is enhancement associated with the $1/r^{3/4}$ growth from $r_i$ down
to the horizon.  The flux in regime IV has an intuitive form when expressed in terms of the virial velocity $v_{\rm typical} = \sqrt{r_s / r_i}$, namely $\Phi = 4 \pi r_i^2 \rho_i v_{\rm typical}.$
That is, the energy density $\rho_i$ flows across the sphere of impact with velocity $v_{\rm typical}$.

\section{Outside the sphere of impact ($r>r_i$)} \label{sec:galaxy}
In the previous section we examined the behavior of a scalar field in a Schwarzschild geometry.  This is a reasonable approximation close to the black hole, but breaks down around the
sphere of impact where the potential of the surrounding matter takes over.  Here we take steps toward modeling a more realistic situation.  We will slightly modify the geometry
far from the black hole in a way that makes the effective potential constant.  This should be viewed as a tractable toy model of a more realistic situation.  The toy model
has the advantage of giving us incoming and outgoing spherical waves.  This makes the analysis and interpretation clean and
allows us to connect the black hole hair problem to a standard scattering problem.

Below we first introduce a toy model for the geometry at $r > r_i$.  Then we compute the transmission and reflection coefficients for an incoming wave, discuss the scalar field profile, and finally consider the energy accretion rate.  We will denote the typical virial velocity $v = \sqrt{r_s/r_i}$.  Our main goal will be to understand the dependence on the scalar mass $m$.  As we vary this mass we will hold $v$ fixed, which means the momentum far from the black hole $k = mv$ will scale with $m$.

\subsection{Modified metric and matching}
To modify the metric we choose a radius $r_i$ and match the Schwarzschild metric for $r < r_i$ to a constant metric for $r > r_i$.  That is, we take
\begin{equation}
\label{ModifiedMetric}
ds^2 = \left\lbrace\begin{array}{ll}
- \left(1 - {1 \over r}\right) dt^2 + \left(1 - {1 \over r}\right)^{-1} dr^2 + r^2 d\Omega^2 & \qquad r < r_i \\[5pt]
- f_i \, dt^2 + f_i^{-1} dr^2 + r^2 d\Omega^2 & \qquad r > r_i
\end{array}\right.
\end{equation}
where $f_i = 1 - {1 \over r_i}$.  The metric is continuous across $r_i$.  For $r > r_i$ there is a constant redshift factor and spatial slices have a conical geometry, but the deviation from Minkowski space is small
provided $r_i \gg 1$ so that $f_i \approx 1$.\footnote{Keep in mind that realistically, the metric does not
maintain this conical form as one goes to even larger radii, where
metric fluctuations of the galaxy comes into play. Our simple toy
model is chosen for simplicity rather than realism. It is worth
emphasizing that none of the results at $r < r_i$ are affected by the
choice of the toy model for what happens at $r > r_i$.}
As follows from appendix \ref{app:KG} the tortoise coordinate is
\begin{eqnarray}
&& r_* = \left\lbrace\begin{array}{ll}
r + {r_i \over r_i - 1} + \log {r - 1 \over r_i - 1} & \quad r < r_i \\[2pt]
{r \over f_i} & \quad r > r_i
\end{array}\right.
\end{eqnarray}
and the effective potential is
\begin{eqnarray}
&& V = \left\lbrace\begin{array}{ll}
m^2 - {m^2 \over r} + {1 \over r^3} - {1 \over r^4} & \quad r < r_i \\[2pt]
m^2 - {m^2 \over r_i} & \quad r > r_i
\end{array}\right.
\end{eqnarray}
Note that the effective potential is not quite continuous across $r_i$.  It has a step discontinuity at $r = r_i$, but the step is small and will not have much effect provided $r_i \gg 1$.

In the Schwarzschild region $r < r_i$ we adopt the solution given in \eqref{smooth}, which is smooth across the future horizon.
On the other hand, for $r > r_i$ we have oscillating solutions,
\begin{equation}
\phi(t,r) = e^{-i \omega t} {1 \over r} R(r) \quad {\rm with} \quad R(r) = b_1 e^{i k r_*} - b_2 e^{-i k r_*} \, .
\label{sol000}
\end{equation}
The relative minus sign is conventional for s-wave scattering.
In \eqref{sol000} the momentum $k$ is real and positive, and it is
related to $\bar{k}$, the momentum at infinity in a pure Schwarzschild
geometry, by\footnote{Note that in general $\bar{k}^2$ can be
  negative as discussed below (\ref{Eeqn}).}
\begin{equation}
k^2 = \bar{k}^2 + m^2 / r_i \, .
\end{equation}
  The  solutions \eqref{sol000} correspond to spherical waves,
\begin{equation}
\label{spherical}
\phi(t,r) = e^{-i \omega t} {1 \over r} \left( b_1 e^{i k r_*} - b_2 e^{-i k r_*} \right) \, ,
\end{equation}
where $b_2$ and $b_1$ are  the amplitudes of the incoming and reflected waves respectively.

Patching the solutions \eqref{smooth} and \eqref{spherical} together by requiring that $\phi$ and $\partial_{r_*} \phi$ are continuous across $r_i$  leads to the system of equations
\begin{align}
&c_2 (r_i - 1)^{-i\omega}e^{-i \bar{k} r_i} r_i {\sf HeunC} = b_1 e^{i k r_{*i}} - b_2 e^{-i k r_{*i}} \, , 
\label{mat1}\\
&c_2 (r_i - 1)^{-i\omega}e^{-i \bar{k} r_i} f_i \left[\left(1 - {i \omega \over f_i} - i \bar{k} r_i\right) {\sf HeunC} - r_i {\sf HeunC}'\right] = i k \left(b_1 e^{i k r_{*i}} + b_2 e^{-i k r_{*i}}\right) \, ,
\label{mat2}
\end{align}
where ${\sf HeunC} = {\sf HeunC}(2i\bar{k}, -2i\omega, 0, -\omega^2 -\bar{k}^2, \omega^2 + \bar{k}^2, 1 - r_i)$ and ${\sf HeunC}'$ denotes a derivative with respect to the last argument.

The matching equations \eqref{mat1}-\eqref{mat2} are somewhat unwieldy but they simplify in various limits.
The most interesting situation is a field that oscillates with frequency $\omega = m$,
which corresponds to setting $\bar{k} = 0$ and $k = m / \sqrt{r_i}$.  Then for small mass and large $r_i$ we can use the expansion (\ref{HCsmall}) in App. \ref{appendix:asymptotics} to find\footnote{Restoring the Schwarzschild radius $r_s$, we are expanding in powers of $m$ and taking $r_s \ll r_i \ll 1/m^2 r_s$.  The last condition is necessary for the validity of the $m$ expansion.}
\begin{eqnarray}
\label{SmallMass}
&& {b_1 \over b_2} = 1 - {2 m^2 \over \sqrt{r_i}} + {i \over 3} r_i^{3/2} m^3 + {\cal O}(m^4) \, , \\
&& {c_2 \over b_2} = {2 i m \over \sqrt{r_i}} - {m^2 \over r_i^{5/2}} + i \sqrt{r_i} m^3 + {\cal O}(m^4) \, ,
\label{SmallMassbis}
\end{eqnarray}
On the other hand, expanding for large mass and large $r_i$ using (\ref{HCbig}) gives
\begin{eqnarray}
\label{LargeMass}
&& {b_1 \over b_2} = {1 \over 2 \sqrt{r_i}} e^{-2im\sqrt{r_i}} \, , \\
&& {c_2 \over b_2} = - {1\over r_i^{1/4}} e^{im\sqrt{r_i}} \, .
\label{LargeMassbis}
\end{eqnarray}

\subsection{Transmission and reflection coefficients}
\begin{figure}
\centerline{\includegraphics[width=9cm]{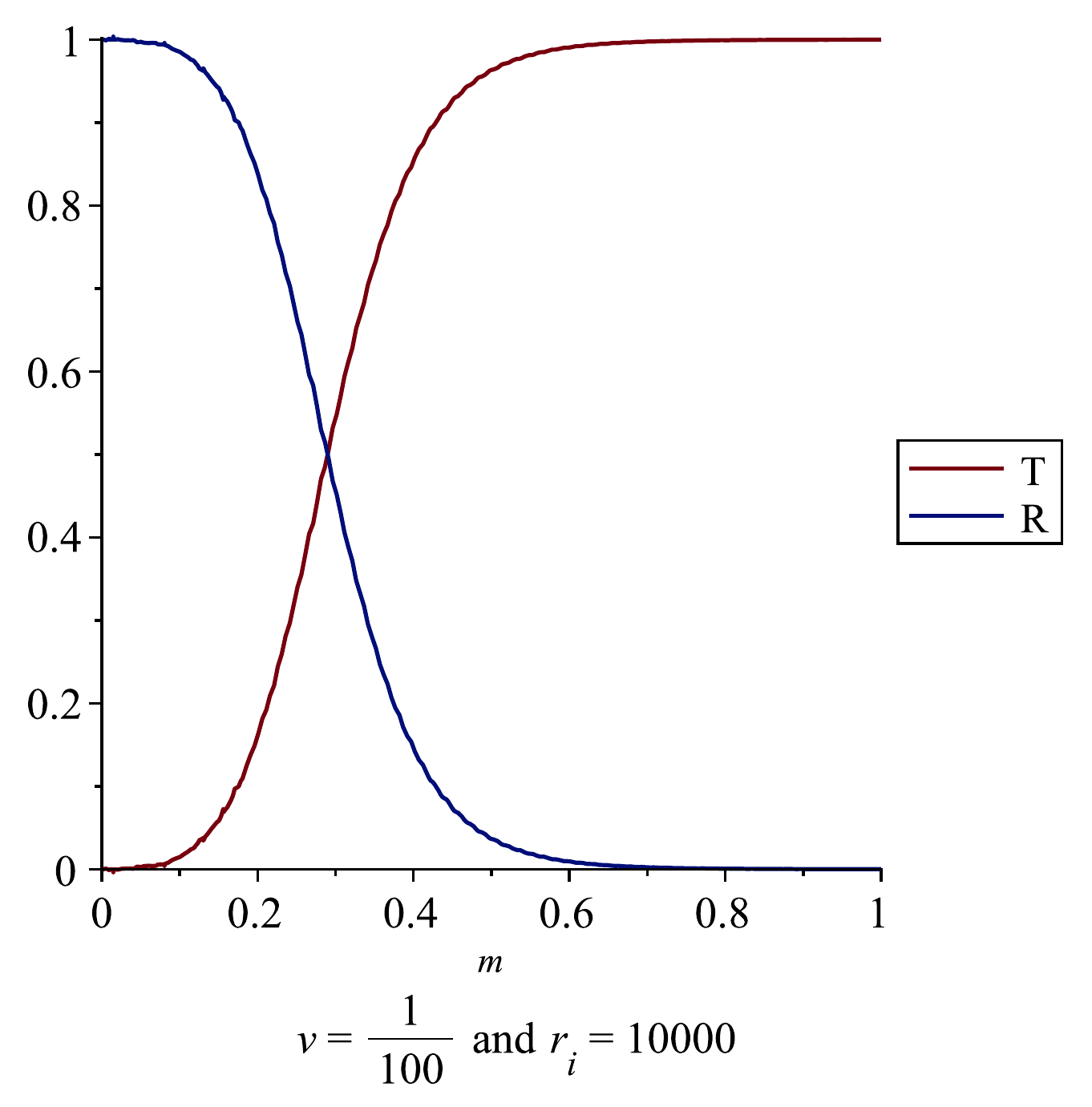}}
\caption{Transmission and reflection coefficients as a function of $m$.\label{fig:TR}}
\end{figure}

We begin by computing the transmission and reflection coefficients for scattering off the black hole.  The energy flux across the horizon was given in \eqref{HorizonFlux0}.
Then, the incoming and outgoing fluxes associated with \eqref{spherical} are
\begin{eqnarray}
&& \Phi_{\rm in} = 8 \pi k \omega \vert b_2 \vert^2 \, , \\
&& \Phi_{\rm out} = 8 \pi k \omega \vert b_1 \vert^2 \, .
\end{eqnarray}
Thus, the transmission and reflection coefficients take on the form
\begin{equation}
{\cal T} = {\omega \over k} \left\vert {c_2 e^{-i \bar{k}} \over b_2} \right\vert^2 \, , \qquad
{\cal R} = \left\vert {b_1 \over b_2} \right\vert^2 \, ,
\end{equation}
with ${\cal T} + {\cal R} = 1$.  These are plotted in Fig.\ \ref{fig:TR} as a function of $m$.  As expected, for small mass the wave mostly reflects off the barrier while for large mass it mostly falls into the black hole.  This also follows from
the expansions \eqref{SmallMass}-\eqref{LargeMassbis}, which imply,  for small mass,
\begin{equation}
{\cal T} \approx {4 m^2 \over \sqrt{r_i}} \, , \qquad {\cal R} \approx 1 - {4 m^2 \over \sqrt{r_i}} \, ,
\end{equation}
while, for large mass,
\begin{equation}
{\cal T} \approx 1\, , \qquad {\cal R} \approx 0\,.
\end{equation}

\subsection{Field profile}

In section \ref{sec:schwarzschild} we explored the field profile near the black hole in various mass regimes.  Here we match those results to the spherical waves (\ref{spherical}) and
discuss the behavior of the field at $r > r_i$. To this  end, we will set $\bar{k} \approx 0$ so that $\omega \approx m$, and we will take $r_i \gg 1$ in such a way  that $f_i \approx 1$.
Note that this means $r_* \approx r$ for $r > r_i$.
There are basically four situations that occur, which, in order of increasing mass, are given by the following.
\begin{enumerate}[label=\Roman*.]
\item
In regime I, $m \lesssim 1/r_i^2$ or equivalently $m \lesssim v^4$, the field for $r < r_i$ is given by retaining the first-order term in (\ref{smallmExpand}),
\begin{equation}
\phi(t,r<r_i) = \sqrt{\rho_i \over 2m^2} e^{-imt} \left(1 - im\log(1 - 1/r)\right)
\end{equation}
where we have normalized as in (\ref{smallm2}).  Since the mass is small, i.e.\ $kr_i = m v r_i \ll 1$, the field does not oscillate on length scales of interest and we can expand the exponentials in \eqref{spherical}
to first order.  Matching the field and its derivative at $r_i$ gives
\begin{equation}
\label{phiI}
\phi(t,r_i < r < 1/k) = \sqrt{\rho_i \over 2 m^2} e^{-i m t} \left(1 + {im \over r}\right) .
\end{equation}
Note that the homogeneous term dominates and the $1/r$ hair is a small correction.  This result also follows from using the small-mass expansion \eqref{SmallMass} to first nontrivial order.
\item
In regime II the mass is larger, $m > v^4$ but $k r_i = m v r_i < 1$.  In this regime we can still expand in powers of the mass.  But in (\ref{smallmExpand}) the second-order term is enhanced by a factor of
$m r^2$ compared to the first-order term and can make the dominant correction to the field profile.  This means we should retain subleading terms in the small-mass expansion (\ref{SmallMass}) -- (\ref{SmallMassbis}) and we should expand the exponentials in (\ref{spherical}) to second order.  Carrying out the matching of the
field and its derivative at $r_i$ gives
\begin{eqnarray}
\label{rIIsmallr}
\phi(t,1 < r < r_i) & = & c_2 e^{-imt} \left(1 + {1 \over 2} m^2 r_i + {i m \over r} - {1 \over 2} m^2 r + {\cal O}(m^3)\right) \\
\label{rIIbigr}
\phi(t,r_i < r < 1/k) & = & c_2 e^{-imt} \left(1 + {1 \over r}\big(im + {1 \over 6} m^2 r_i^2\big) - {m^2 r^2 \over 6 r_i} + {\cal O}(m^3)\right)
\end{eqnarray}
where $c_2 \approx \sqrt{\rho_i / 2 m^2}$.

In (\ref{rIIsmallr}) note that the field amplitude near the horizon is enhanced by a factor $1 + m^2 r_i / 2$ (an effect which was neglected in (\ref{smallm1})).
Also the terms quadratic in the mass, although they are still small compared to 1 since $kr_i = mvr_i = m \sqrt{r_i} < 1$, can dominate over the term linear in the mass.\footnote{In appendix \ref{appendix:asymptotics} this corresponds to the fact that the expansion in powers of $m$ involves both $mr_s$ and $m\sqrt{r r_s}$.}  For the term linear in $m$ to dominate all the
way out to $r = r_i$ we require $m r_i^2 < 1$ or equivalently $m r_s < v^4$, which would put us back in regime I.  In (\ref{rIIbigr}) the quadratic $m^2 r^2$ term can be understood as the expansion of a standing wave
$\sim {1 \over kr} \sin kr$ to quadratic order in $k$.  When $m r_s > v^4$ the standing wave starts to dominate over the $i m / r$ tail in the field profile.  But the $i m / r$ tail is still important
because the standing wave carries no flux.  So energy transport into the black hole still comes from the $i m / r$ tail.
\item
As the mass increases further we enter a regime where $k r_i = m v r_i > 1$ and the field begins to oscillate.  With the scaling $v^2 \sim 1/r_i$ this happens when $m \gtrsim v$, so we are not yet in the regime $m \gtrsim 1$ where the wave
can easily fall into the black hole.  Instead, it reflects off the potential and sets up a standing wave outside the black hole.
Matching the solution (\ref{RegimeIIIsol}) to spherical waves at $r > r_i$ gives
\begin{eqnarray}
\phi(t,1/m^2 < r < r_i) & = & \sqrt{\rho_i \over m^2} \left({r_i \over r}\right)^{3/4} e^{-imt} \cos\big(2m\sqrt{r} - 3\pi/4\big) , \\
\phi(t,r > r_i) & = & \sqrt{\rho_i \over m^2} {r_i \over r} e^{-imt} \cos\big(kr + m\sqrt{r_i} - 3\pi/4\big) .
\end{eqnarray}
This matching makes the field and its first derivative continuous across $r_i$.  To see this it is enough to recall that $\omega \approx m$ implies $k \approx m / \sqrt{r_i}$.
\item
In regime IV, $m > 1$, the barrier around the black hole disappears and the wave is purely ingoing.  Matching (\ref{RegimeIVsol}) to an ingoing spherical wave gives
\begin{eqnarray}
\phi(t,1 < r < r_i) & = & \sqrt{\rho_i \over 2 m^2} \left({r_i \over r}\right)^{3/4} e^{-imt} e^{-2im\sqrt{r}} ,\\
\phi(t,r > r_i) & = & \sqrt{\rho_i \over 2 m^2} {r_i \over r} e^{-imt} e^{-i(kr + m\sqrt{r_i})} .
\end{eqnarray}
\end{enumerate}

\begin{figure}
\centerline{\includegraphics[width=9cm]{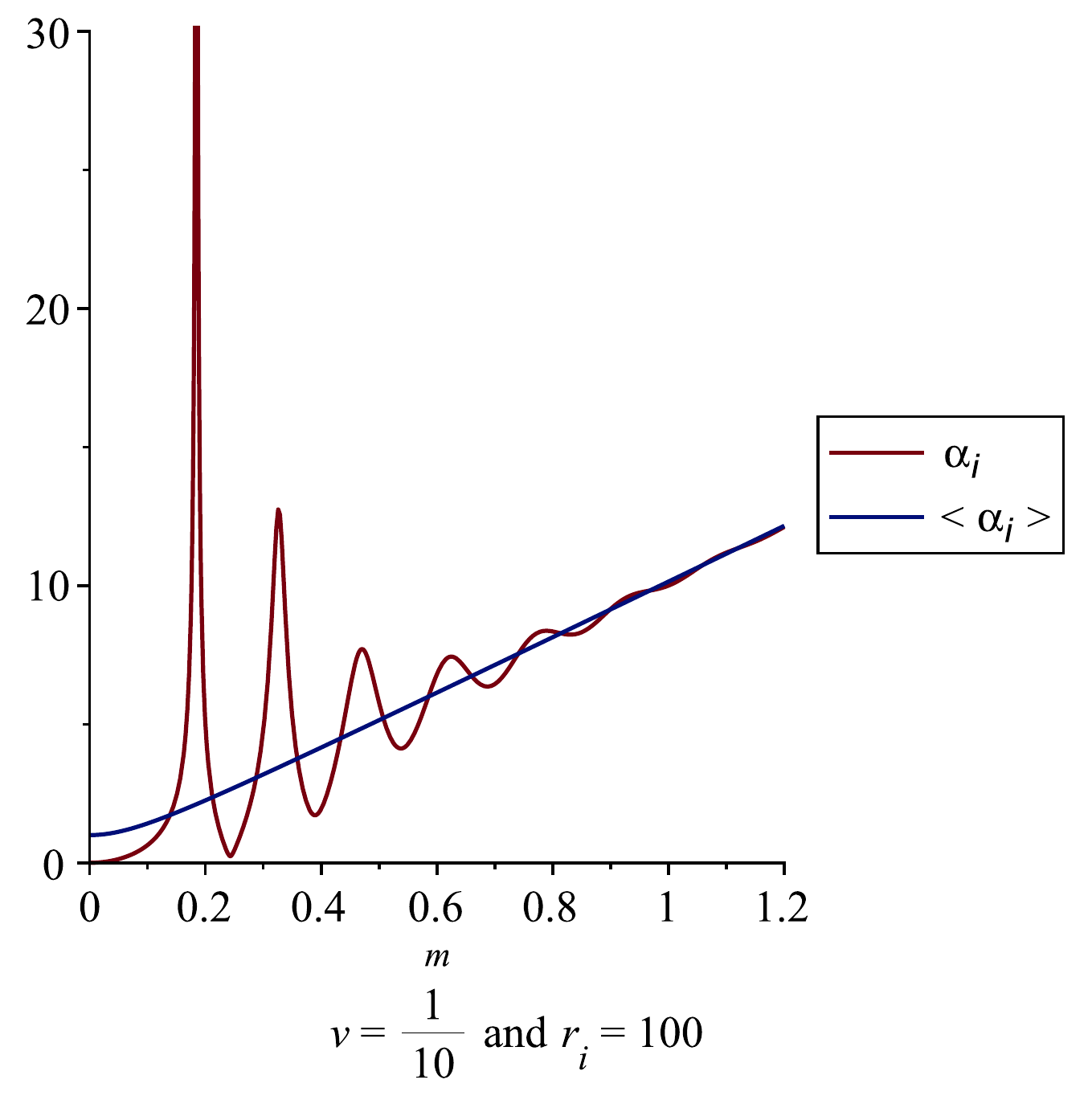}}
\caption{$\alpha_i$ and $\langle \alpha_i \rangle$ as a function of $m$.\label{fig:alpha}}
\end{figure}

As a quantity that can help us distinguish between these different behaviors we introduce
\begin{equation}
\alpha_i = {\vert r \partial_r \phi\vert_{r=r_i} \over \vert\phi\vert_{r=r_i}} \, .
\end{equation}
Thus,  from the discussion above:
\begin{eqnarray*}
&&\hbox{\rm I,\,II. \quad \!\!\! for low masses, we expect $\alpha_i \approx 0$, reflecting a homogeneous field;} \\
&&\hbox{\rm III. \quad once a standing wave develops, the field has nodes so $\alpha_i$ should have spikes;} \\
&&\hbox{\rm IV. \quad once the wave is purely ingoing the spikes go away.}
\end{eqnarray*}
As a related quantity, we can spatially average $\alpha_i$ over an oscillation.\footnote{To illustrate the averaging procedure, for $\phi$ in (\ref{spherical}) we would define
$\langle \vert \phi \vert^2 \rangle = {1 \over r^2} \left(\vert b_1 \vert^2 + \vert b_2 \vert^2\right)$.}  This leads to
\begin{equation}
\langle \alpha_i \rangle \equiv \sqrt{\langle \vert r \partial_r \phi\vert^2 \rangle \over \langle \vert \phi \vert^2 \rangle} = \sqrt{1 + {k^2 r_i^2 \over f_i^2}} \, .
\end{equation}
The averaging smooths out the spikes, but it also throws out the phase coherence responsible for the homogeneous field at low mass.  So at low mass we expect $\alpha_i \rightarrow 0$ but $\langle \alpha_i \rangle \rightarrow 1$.
These behaviors can be seen in Fig.\ \ref{fig:alpha}.

\subsection{Energy accretion rate\label{sect:accretion}}
Finally we study the rate at which the black hole gains energy. In particular, we would like to compare  the energy density in the scalar field at $r_i$,
\begin{equation}
\rho_i \equiv T^{tt} = {k^2 + 2 m^2 f_i \over f_i^2} \vert \phi \vert^2 + \vert \partial_r \phi \vert^2  \qquad \hbox{\rm evaluated at $r = r_i$} \, ,
\end{equation}
with the flux of energy $\Phi_{\rm horizon}$ entering the horizon---see Eq. \eqref{HorizonFlux0}.
To this end, we introduce a quantity ${\cal V}$, defined by
\begin{equation}
\Phi_{\rm horizon} = 4 \pi r_i^2 \, \rho_i \, {\cal V} \, .
\end{equation}
Since ${\rm flux} = {\rm area} \times {\rm density} \times {\rm velocity}$, we can interpret ${\cal V}$ as the velocity at which the energy in the scalar field crosses $r_i$.
As in the previous subsection, it is useful to average over spatial oscillations.  This leads to the averaged energy density
\begin{equation}
\langle \rho_i \rangle = {1 \over f_i^2} \Big(2 m^2 f_i + 2k^2 + {f_i^2 \over r_i^2}\Big) {1 \over r_i^2} \Big(\vert b_1 \vert^2 + \vert b_2 \vert^2\Big) \, ,
\end{equation}
which we can relate to the averaged velocity $\langle {\cal V} \rangle$ by
\begin{equation}
\Phi_{\rm horizon} = 4 \pi r_i^2 \, \langle \rho_i \rangle \, \langle {\cal V} \rangle \, .
\label{flux0}
\end{equation}

\begin{figure}
\centerline{\includegraphics[height=9cm]{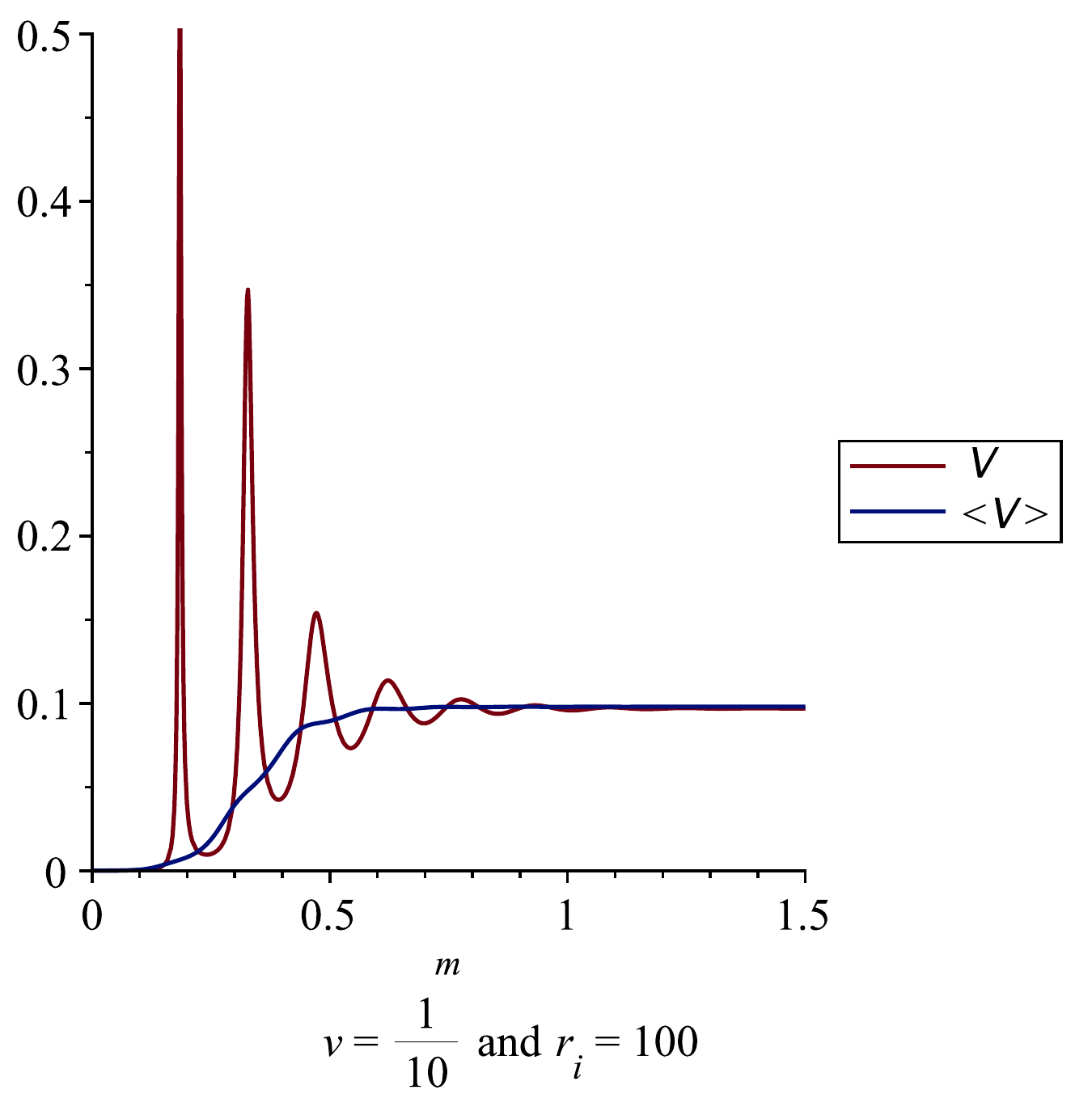} \includegraphics[height=9cm]{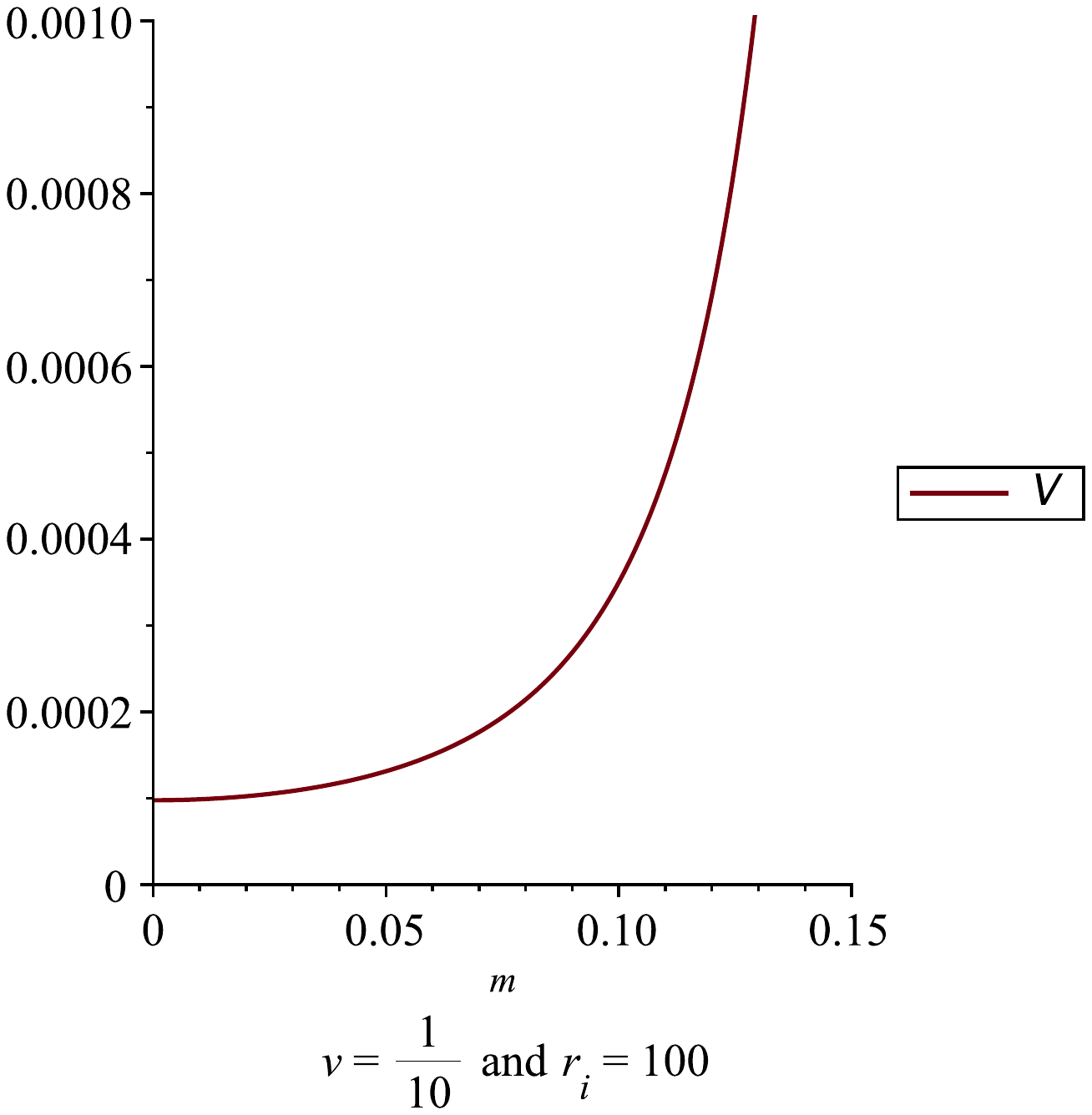}}
\caption{${\cal V}$ and $\langle {\cal V} \rangle$ as a function of $m$.  The right panel zooms in on small mass.  For small mass phase coherence is important so we don't show $\langle {\cal V} \rangle$.\label{fig:V}}
\end{figure}

We  plot these quantities in Fig.\ \ref{fig:V}, where we set $\omega = m$ and $r_i \gg 1$.  As expected, averaging smooths out the spikes in ${\cal V}$.  As can be seen in the figure, at large mass values  ${\cal V} \rightarrow v$, while for small masses ${\cal V} \rightarrow 1/r_i^2$.
These results also follow from the expansions \eqref{SmallMass}-\eqref{LargeMassbis}.  Thus restoring the Schwarzschild radius $r_s$ and the speed of light $c$ we have
\begin{eqnarray}
&&\hbox{\rm large mass:} \quad \Phi_{\rm horizon} = 4 \pi r_i^2 \, \rho_i \, v  \, ,\\
\label{SmallMassFlux}
&&\hbox{\rm small mass:} \quad \Phi_{\rm horizon} = 4 \pi r_s^2 \, \rho_i \, c \, ,
\end{eqnarray}
in agreement with  (\ref{PhiOneTwo}) and (\ref{PhiFour}).  The first line is hardly a surprise---it is simply saying that energy moves across $r_i$ with the velocity of the dark matter particles.  To gain some intuition about the second line it is useful to switch to tortoise coordinates.  In tortoise coordinates near the horizon the field is effectively massless and everything moves
at the speed of light.  The energy density in the tortoise coordinate is
\begin{equation}
\rho_* = \rho {\D r \over \D r_*} = f^2 T^{tt} 
\end{equation}
which close to the horizon becomes $\rho_* \approx 2 m^2 \vert \phi \vert^2$.  On the other hand for large $r_i$ we have $\rho_i \approx 2 m^2 \vert \phi \vert^2$.  For small mass, somewhat
surprisingly, the field amplitude is the same near the horizon and at $r_i$.  So $\rho_i$ can be identified with the near-horizon energy density in tortoise coordinates,
which accounts for (\ref{SmallMassFlux}).

\section{Discussion}
\label{sec:conclude}

In this paper, we generalize a phenomenon first noted by Jacobson
\cite{Jacobson:1999vr}, that a time-dependent boundary condition for a
scalar can endow a black hole with scalar hair. Our set-up is
motivated by (1) the fact that any astrophysical black hole is surrounded
by dark matter, and (2) the notion that dark matter might be a scalar field, whose
non-zero mass implies inevitable oscillations in time.
The question is how much hair would be generated around the black
hole.

To address this question, we revisit the Klein Gordon equation in a
Schwarzschild background. This is an old subject with a vast
literature (see e.g. \cite{Unruh:1976fm,Detweiler:1980uk} for early
papers). Our goal is to fill, as far as we know, a certain gap in the
literature, working out the scalar profile around the black hole, as the scalar
mass is systematically varied. Three assumptions are made in our
computation:
(1) that the angular momentum of the scalar is ignored,
as is the spin of the black hole,
(2) that the gravitational backreaction of the scalar is
negligible, and (3) that the possible self-interaction of the scalar---for
instance, if it were an axion---is unimportant.

\begin{table}[tb]
\begin{center}
\begin{tabular}{ccl}
regime & mass & \hspace{4mm} $\phi(r_s \lesssim r < r_i)$ \\
\hline
I & $mr_s < v^4$ & \hspace{2mm} $ \sqrt{\rho_i \over 2 m^2} \left(1 + {im r_s^2 \over r} \right)$ \rule{0pt}{16pt}\\
II & $ v^4 < mr_s < v $ & \hspace{2mm} $\sqrt{\rho_i \over 2 m^2} \left(1 + {im r_s^2  \over r} - {1\over 2} m^2 r r_s  \right)$ \rule{0pt}{16pt}\\
III & $v< mr_s < 1$ & $\left\lbrace
\begin{array}{ll}
\sqrt{\pi \rho_i \over m^2} (m^2 r_i r_s)^{3/4} & \quad r_s \lesssim r \lesssim 1/m^2 r_s \\[4pt]
\sqrt{\rho_i \over m^2} \left( {r_i \over r}  \right)^{3/4} {\,\rm cos\,} \left(2m \sqrt{r r_s} - 3\pi/4\right) & \quad 1/m^2 r_s \lesssim r < r_i
\end{array}\right.$ \rule{0pt}{28pt}\\
IV & $mr_s > 1$ & \hspace{2mm} $\sqrt{\rho_i \over 2 m^2} \left( {r_i \over r}  \right)^{3/4}  e^{-i 2 m\sqrt{r r_s}}$ \rule{0pt}{18pt}
\end{tabular}
\end{center}
\caption{Field profile in different mass ranges.  We set $\omega = m$ and denote the virial velocity by $v = \sqrt{r_s / r_i}$.
For simplicity time dependence and some subleading terms have been suppressed.
\label{table:field}}
\end{table}

We will address each of these points below, but let us first briefly summarize
our findings. The scalar field has distinct profiles depending on the
size of scalar mass $m$. There are four regimes, summarized in
Fig. \ref{Fig:1} and Table \ref{table:field}. Besides the scalar mass
$m$, there are two additional scales in the problem: $r_s$ the
Schwarzschild radius of the black hole of course, and $r_i$ the radius
of sphere of impact, meaning at distances within $r_i$ the geometry
is dominated by the black hole. It is helpful to define a velocity
scale $v \equiv \sqrt{r_s / r_i}$ which is the typical velocity
dispersion at $r_i$. Regime I, the extreme low mass limit (or wave
limit), is one
where $m r_s < v^4$. The scalar field more or less oscillates with
the same amplitude, from $r_i$ all the way down to $r_s$.
There is a $1/r$ component, whose coefficient can be identified
as the scalar charge of the black hole, but it is very small.
Regime II, where $v^4 < m r_s < v$, is very similar, with an additional
(again small) linear $r$ component. 
Regime III, with $v < m r_s < 1$, is perhaps the most interesting
regime where one sees both particle-like and wave-like behavior.
The scalar has a $1/r^{3/4}$ profile at large $r$ which is
characteristic of a particle (to be elaborated below), 
but also has a ${\,\rm cos\,} (2m \sqrt{r r_s} - 3\pi/4)$ modulation,
characteristic of a standing wave.
Regime IV, with $m r_s > 1$ i.e. the scalar Compton wavelength is
smaller than the horizon size, is the particle limit with a
$1/r^{3/4}$ profile but without the wave-like modulation.
Observationally, it is probably challenging to measure the dark matter
profile around an astrophysical black hole, but if it were possible,
the profile can be used to deduce the dark matter mass.
The standing wave nodes in regime III are particularly distinctive
features to look for.

How do we understand the $1/r^{3/4}$ profile in the particle limit? 
Imagine a particle falling towards the black hole from far away,
it acquires a speed $u$ that goes as $1/r^{1/2}$ by energy
conservation.\footnote{For this argument, it does not matter a whole
  lot what precise velocity the particle originally had far away from the
  black hole. After some free fall towards the black hole, the
  particle's velocity would be dominated by the one generated by
  gravity, thus going as $1/r^{1/2}$.}
In our computation, we are essentially looking for a stationary
configuration of particles accreting onto the black hole.
Such a stationary configuration would have an infalling flux that is
independent of radius. In other words, we expect $4\pi r^2 \rho u$ to be
independent of $r$. Thus, $\rho \propto 1/r^{3/2}$. Since
$\rho \sim m^2 \phi^2$, the scalar field $\phi$ has a profile
of $1/r^{3/4}$. This argument also explains the phase of $\phi$ in regime
IV (see Eq. (\ref{equation:RegimeIV})): differentiating it with
respect to $r$ gives the particle momentum, and indeed the resulting
$1/r^{1/2}$ is consistent with the velocity $u$ we deduce by this
simple argument. Phrased in this way, our black hole hair is in a
sense fairly mundane: it is 
nothing other than a steady accretion flow of
  matter onto the black hole. The form of the flow changes as one
  dials the mass, from the particle limit to the wave limit.

It is worth noting that this argument is essentially what went into
earlier discussions about a possible dark matter spike close to the 
black hole at the Galactic center, where the dark matter is assumed to be a heavy particle
\cite{Gondolo:1999ef} (the density profile of $1/r^{3/2}$ can be seen
in that context as originating from an initially flat one). Subsequent
authors pointed out that the dark matter spike can be destroyed
by dynamical processes as the seed black hole spiral to the center by
dynamical friction \cite{Ullio:2001fb}. The same caveats apply to
our idealized computation as well, though for a sufficiently small
dark matter
particle mass, the resulting soliton that typically condenses at the
center of galaxies \cite{Schive:2014hza} could have a stabilizing effect, a point we will
come back to below.

Let us turn to the three assumptions outlined earlier.
First, our computation is done largely assuming s-wave, 
or $l = 0$ (except in appendix \ref{app:KG}). 
This is a good approximation in the small mass limit.
The angular momentum of a particle can be estimated as
$m  v  r_i$, where $v$ is defined earlier---the typical
virial velocity at the radius of sphere of impact $r_i$ 
i.e. $v = \sqrt{r_s / r_i}$. Thus, $m v r_i = m r_s / v$ which
is less than unity as long as we are in regime I or II. 
For regime III and IV, ignoring angular momentum is no longer
justified. However, at sufficiently large $r$---basically outside the
angular momentum barrier which produces a turning point at $r \sim l^2 / (m^2 r_s)$---the $1/r^{3/4}$ profile remains valid
(see appendix \ref{app:KG} for justification). 
Inside the angular momentum barrier, the scalar profile would depend
on the angular momentum distribution of the scalar.
There is of course also the angular momentum of the black hole itself,
which we have ignored. For a spinning black hole, we expect our
results to be applicable at a sufficiently large $r$, but there could be
non-trivial effects, instability even, for instance the well known
superradiance effect---we will return to this point below.

A second assumption we make is the absence of gravitational
backreaction from the scalar. We can estimate the curvature produced
by the dark matter scalar divided by the curvature sourced by the black hole, at
around the horizon, by $16 \pi G r_s^2 \rho$. Here $\rho$ is the dark
matter density close to the black hole, which is at best $\rho_i
(r_i/r_s)^{3/2}$ (the high mass limit), with $\rho_i$ being the dark
matter density at $r_i$ far away from the black hole. Therefore:
\begin{equation}
\label{r2}
16 \pi G r_s^2 \rho \lesssim 16\pi G r_s^2 \rho_i
\left(\frac{r_i}{r_s} \right)^{3/2}
\sim 6 \times 10^{-13} \left( {M_{\rm
        BH} \over 10^9 {\,\rm M_\odot}}\right)^2 \left({\rho_i \over 1 {\,\rm GeV\,}/{\,\rm
      cm^3\,}}\right) \left({r_i/r_s \over 10^6}\right)^{3/2} \, .
\end{equation}
The value of $\rho_i \sim 1 {\,\rm GeV\,}/{\,\rm cm^3}$ is about the dark matter
density in the solar neighborhood. The gravitational backreaction is
weak. Even boosting $\rho_i$ by a few orders of
magnitude for a black hole in denser parts of the galaxy (or in denser
galaxies) would not
alter the basic smallness of the effect.
See also \cite{Barausse:2014tra} for estimates of related environmental effects around black hole binaries.
A corollary of the weak gravitational backreaction is that the scalar hair
would also be difficult to observe, since our primary way of deducing
the existence of dark matter is via its gravitational effect.

A third assumption we make is that the scalar has negligible
self-interaction. An appealing scalar dark matter candidate is the
axion, or axion-like-particle. The axion is expected to have
self-interaction, for instance a quartic term in the Lagrangian of the form ${\cal L}_{\rm
  quartic} \sim (m/F)^2 \phi^4$, where $F$ is the axion decay
constant. One can estimate the importance of the self-interaction by
taking the ratio of the quadratic mass term $\sim m^2 \phi^2$ and the
quartic term, i.e. $\phi^2 / F^2 \sim \rho / (m^2 F^2)$. 
Using the same reasoning as before, this is at best:
\begin{eqnarray}
&& {\rho \over m^2 F^2} \lesssim \rho_i \left(\frac{r_i}{r_s}
  \right)^{3/2} {1\over m^{2} F^{2}} \nonumber \\
&& \quad \quad \quad \quad \sim 8 \times 10^{-7}
\left({\rho_i \over 1 {\,\rm GeV\,}/{\,\rm
      cm^3\,}}\right) \left({r_i/r_s \over 10^6}\right)^{3/2}
\left( {10^{-21} {\,\rm eV} \over m} \right)^2
\left( {10^{17} {\,\rm GeV} \over F} \right)^2 .
\end{eqnarray}
Note that the upper limit makes use of the pile-up of the
scalar close to the horizon, which applies only if the scalar
mass is sufficiently large i.e. at least in regime III if not in
regime IV i.e. 
$m \, \gsim \, 7 \times 10^{-23} {\,\rm eV} (10^9 M_\odot / M_{\rm BH}) 
(10^6 / (r_i/r_s))^{1/2}$. It appears the self-interaction associated
with an axion is also small, though the size depends the values for
$m$ and $F$. For the values chosen, the self-interaction effect is in fact a
lot larger than the gravitational backreaction.

The above discussion suggests an interesting case to consider
is one where the scalar mass is sufficiently small
that a soliton is expected to condense at the center of halos \cite{Schive:2014hza},
a possibility often referred to as fuzzy dark matter
\cite{Hu:2000ke}. In such a case, the corresponding $\rho_i$ can be
much larger, {\it if} the black hole resides in the soliton.
Let us investigate this in a concrete example:
the supermassive black hole 
$M_{\rm BH} \sim 6.5 \times 10^9 M_\odot$
in M87, recently imaged by the Event
Horizon Telescope
\cite{Akiyama:2019cqa,Akiyama:2019eap,Akiyama:2019fyp}.
Its horizon size is equal to the Compton wavelength of a particle
of mass $\sim 10^{-20} {\,\rm eV}$. From numerical simulations, 
the soliton mass is related to the halo mass by \cite{Schive:2014hza}:
\begin{eqnarray}
\label{solitonhalo}
M_{\rm soliton} \sim 2 \times 10^9 M_\odot 
\left( {10^{-22} {\, \rm eV} \over m} \right) \left( {M_{\rm halo} \over 2
  \times 10^{14} M_\odot} \right)^{1/3} ,
\end{eqnarray}
where we adopt the mass of the Virgo cluster halo in which M87
resides \cite{Hui:2016ltb}. The corresponding soliton radius is:
\begin{eqnarray}
\label{solitonsize}
R_{\rm soliton} \sim 5 \times 10^{15} {\,\rm km\,} 
\left( {2 \times 10^9 M_\odot \over M_{\rm soliton}} \right)
\left( {10^{-22} {\,\rm eV\,} \over m} \right)^2 \, .
\end{eqnarray}
Using the results for the scalar pile-up in regime III
and IV, we find that
around the horizon, the dimensionless measure of the importance of
self-interaction $\phi^2 / F^2 \sim \rho / (m^2 F^2)$ is 
$\sim
10^{-7} - 10^{-3}$ for $m \sim 10^{-22} {\rm \, eV\,} 
- 10^{-20} {\rm \, eV\,}$.\footnote{For these estimates, we use the soliton density for $\rho_i$
  and the soliton radius for $r_i$. Note that
the soliton size and density should be altered by the presence of the
black hole itself. Our estimates of the gravitational backreaction and
the importance of self-interaction, based on the unaltered soliton
size and density, are conservative: they are likely underestimates by
about an order of magnitude. We thank Ben Church and Jerry Ostriker
for discussions on this point.
}
These numbers, while small, are potentially important for
superradiance considerations, appropriate in the context of a scalar
Compton wavelength matching roughly the horizon size.\footnote{See \cite{Davoudiasl:2019nlo} for a recent discussion of constraints from
the superradiance argument applied to the M87 black hole. For this
argument, it is important to know the black hole spin, which 
is deduced primarily from the existence of a
jet, rather than from the Event Horizon Telescope data.
One main source of uncertainty on the spin is what fraction of the observed jet
power could be powered by the accretion disk surrounding the black
hole. We thank Eliot Quataert for discussions on this point.
}
Recall that the superradiance
instability is rather slow: the growth rate is at best about $10^{-7}$
times the natural scale in the problem $1/r_s$ \cite{Dolan:2007mj}.
Self-interaction (or gravitational backreaction for that matter) introduces
mixing between the superradiant mode and the scalar hair. 
The effect is weak but might be sufficient to have a
non-negligible impact on superradiance. A detailed discussion is
beyond the scope of this paper---for one thing, our computation needs
to be generalized to a Kerr background---suffice to say the most interesting
mixing comes from a possibly non-axisymmetric component of the hair
(see \cite{Arvanitaki:2010sy} for discussions).
It is also worth emphasizing that the supermassive black hole has the
potential to swallow the soliton, as discussed in \cite{Hui:2016ltb}.
The accretion rate is $\sim 10^{-2} - 10^5 M_\odot / {\,\rm year\,}$
for $m \sim 10^{-22} {\rm \, eV\,} - 10^{-20}
{\rm \, eV\,}$, using the flux appropriate for regime III and IV.
\footnote{The dimensionless number quantifying the importance
of gravitational backreaction, $16\pi G r_s^2 \rho$, is related to the
accretion rate. In all cases, it is negligible: at most $10^{-7}$ for
$m \sim 10^{-20}$ eV.}
There are several caveats, however: (1) the soliton-halo relation
in Eq. (\ref{solitonhalo}) is an extrapolation from simulations of
less massive halos; (2) the soliton might not have enough time to form
(see discussion of relaxation time in \cite{Hui:2016ltb}); (3) the impact of the black hole on
soliton properties should be properly taken into account; (4) the
black hole might not reside in the soliton.  It would be interesting to perform simulations
to map out these possibilities and understand the evolution of the soliton in more detail.

Let us end with a brief discussion of another possible scalar
self-interaction.
Instead of coming from a potential, such as in the case of the axion,
the self-interaction could involve derivatives, such as in the case of
superfluid dark matter \cite{Berezhiani:2015bqa}.
In this case, the Lagrangian is generically a function $P(X)$ of
the kinetic term $X\equiv
-\frac{1}{2}\partial^\mu\phi\partial_\mu\phi$ and it is not hard to find  solutions in the form $\phi(t,r)=t+\psi(r)$ that reduce to
the Jacobson's regular solution \eqref{jacsol} near the horizon. In
particular, for specific choices of the function $P(X)$, the field's
non-linearities can significantly change the scalar profile at
distances of order of the Schwarzschild radius and enhance the
estimate \eqref{r2} by several orders of magnitude. These aspects will
be analyzed in more details in a separate work. 

\acknowledgments

We thank Pedro Ferreira, Austin Joyce, Alberto Nicolis, Riccardo Penco, Rachel Rosen, Enrico Trincherini and Michael
Zlotnikov for discussions. We are grateful to Katy Clough, Pedro Ferreira and Macarena Lagos for sharing their manuscript with us. LH thanks Jerry Ostriker, Scott Tremaine
and Edward Witten for earlier discussions on the black hole accretion
of a scalar.
DK thanks the Columbia University Center for Theoretical Physics for
hospitality during this work.  
The work of DK is supported by U.S.\ National Science Foundation grant PHY-1820734.  
LS is supported by Simons Foundation Award Number 555117.
LH and XL acknowledge support from NASA NXX16AB27G and DOE DE-SC011941. SW is supported in part by the Croucher Foundation. 

\appendix

\section{Scalar wave equation in Schwarzschild geometry}
\label{app:KG}

In this appendix we construct solutions to the  Klein-Gordon equation for a massive scalar field in a Schwarzschild geometry in terms of the confluent Heun function, focusing in particular on solutions  that are smooth across the future horizon.  For generality we first set up the wave equation in a general static spherically-symmetric background, and in obtaining solutions we
include angular momentum.

Given a general static, spherically symmetric background metric,
\begin{equation}
\label{staticapp}
\D s^2 = - f(r) \D t^2 + {1 \over g(r)} \D r^2 + r^2 \D\Omega^2 \, ,
\end{equation}
the ansatz $\phi(t,r, \hat{\Omega})= e^{-i\omega t} \frac{R(r)}{r} Y_{l,n}(\hat{\Omega}) $ allows to rewrite the wave equation
\begin{equation} \label{eqn:scalareom}
- {1 \over f} \partial_t^2 \phi + {1 \over r^2} \Big({g \over f}\Big)^{1/2} \partial_r \Big(r^2 (fg)^{1/2} \partial_r \phi\Big) - m^2 \phi = 0 
\end{equation}
as an effective radial Schr\"odinger equation 
\begin{equation}
\label{radial}
\Big(-\partial_{r_*}^2 + V\Big) R = \omega^2 R \, ,
\end{equation}
where the tortoise coordinate and $R(r)$ are  defined by
\begin{equation}
\D r_* = {\D r \over (fg)^{1/2}},  \quad  \phi(t,r) = e^{-i\omega t} \frac{R(r)}{r} 
\end{equation}
and the potential is
\begin{equation}
\label{SchwarzschildPotential}
V(r) = f \left(m^2 + \frac{l(l+1)}{r^2} \right) + {1 \over 2 r} \partial_r (fg) \, .
\end{equation}
We now specialize to a Schwarzschild geometry and set
\begin{equation}
f(r) = g(r) = 1 - {1 \over r} \, .
\end{equation}
 Thus, the tortoise coordinate becomes $r_* = r + \log \left(r - 1\right)$ and the potential takes on the form
\begin{equation}
\label{Veff}
V(r) = \left(1 - {1 \over r}\right)\left(m^2 + \frac{l(l+1)}{r^2}+  {1 \over r^3}\right) \, .
\end{equation}
The general solution to \eqref{radial} is given by confluent Heun functions \cite{MR1392976,slavjanov2000special},
\begin{multline}
R(r) =  c_1 \, r (r - 1)^{i \omega} e^{i \bar{k} r} \, {\sf HeunC}(-2i\bar{k}, 2i\omega, 0, -\omega^2 -\bar{k}^2, \omega^2 + \bar{k}^2 - l(l+1), 1 - r) 
\\
+ c_2 \, r (r - 1)^{-i\omega} e^{-i \bar{k} r} \, {\sf HeunC}(2i\bar{k}, -2i\omega,0,-\omega^2-\bar{k}^2, \omega^2 + \bar{k}^2 -l(l+1), 1-r) \, .
\label{RconfH}
\end{multline}
Since we are interested in solutions that are smooth across the future horizon, we now examine the near-horizon behavior.  As $r \rightarrow 1$ the confluent Heun functions approach 1, therefore
\begin{equation}
\label{NearHorizon}
\phi(t,r) \approx c_1 \, e^{-i \omega (t - r_*)} e^{-i(\omega - \bar{k})} + c_2 \, e^{-i \omega (t + r_*)} e^{i(\omega - \bar{k})} \, .
\end{equation}
Imposing outgoing boundary conditions, as required by causality,  is equivalent to setting $c_1 = 0$.
This restricts us to the following solution which is smooth across the future horizon:
\begin{equation}
\label{smoothapp}
\phi(t,r) = c_2 \, e^{-i \omega t} (r - 1)^{-i \omega} e^{-i \bar{k} r} \, {\sf HeunC}(2i\bar{k}, -2i\omega, 0, -\omega^2 -\bar{k}^2, \omega^2 + \bar{k}^2-l(l+1), 1 - r) \, .
\end{equation}
In order to systematically study large $r$ asymptotics, we can first write (\ref{radial}) as 
\begin{equation}
  \left( \partial_r ^2 + \bar{k}^2 + \frac{4 \bar{k}^2 + 2m^2 -2l(l+1)-1}{2(r-1)} + \frac{2l(l+1)+1}{2r} + \frac{4\bar{k}^2 + 4m^2 +1}{4 (r-1)^2 } + \frac{1}{4r^2} \right) \left(  R \sqrt{1-\frac{1}{r}}  \right) =0.
\end{equation}
This gives the solution in terms in ${\sf HeunC}$ above. In the large $r$ limit, one can look for asymptotic solution by expanding in $\frac{1}{r-1}$,
\begin{equation}
  \left( \partial_r ^2 + \bar{k}^2 + \frac{2 \bar{k}^2 + m^2}{r-1}  + \frac{\bar{k}^2 + m^2 -l(l+1)}{(r-1)^2 } +{\cal O}\left(\frac{1}{(r-1)^3} \right) \right) \left(  R \sqrt{1-\frac{1}{r}}  \right) =0,
\end{equation}
where all the higher order terms do not depend on $m$ and $k$. At this order the solution is given by the well-studied function ${}_1F_1$. 
\begin{align}  \label{eqn:heuncapprox}
 &R(r) \overset{ r \gg 1 }{\approx}  \tilde{c}_1 \sqrt{r} (r-1)^{\frac{1}{2}\Delta_{\omega,l} } e^{-i \bar{k}r} {}_1F_1 \left( i \bar{k} + i \frac{m^2}{2 \bar{k}} + \frac{1+\Delta_{\omega,l} }{2},1+\Delta_{\omega,l}  , 2 i \bar{k} (r-1) \right) \nonumber \\
 & \quad \qquad+ \tilde{c}_2  \sqrt{r} (r-1)^{-\frac{1}{2}\Delta_{\omega,l} } e^{-i \bar{k}r} {}_1F_1 \left( i \bar{k} + i \frac{m^2}{2 \bar{k}} +  \frac{1-\Delta_{\omega,l} }{2},1-\Delta_{\omega,l} , 2 i \bar{k} (r-1) \right), \nonumber \\
&\Delta_{\omega,l} = i\sqrt{4\omega^2 - 4 l(l+1)-1}.
\end{align}
Therefore, from the asymptotic expansion of ${}_1F_1$ at large $r$, we have 
\begin{equation}
\begin{split}
\phi(t,r)&  \approx  c_3 \, e^{-i \omega t} \, {1 \over r} e^{i \bar{k} r + i (\bar{k} + {m^2 \over 2\bar{k}}) \log (r-1)} (1 + {\cal O}(1/r))
\\
 & \quad +  c_4 \, e^{-i \omega t} \, {1 \over r} e^{-i \bar{k} r -i (\bar{k} + {m^2 \over 2\bar{k}}) \log(r-1)} (1 + {\cal O}(1/r)) \, .
 \end{split}
 \label{asymphiso}
\end{equation}
These are the usual outgoing and ingoing spherical waves of flat space, with a logarithmic distortion due to the Newtonian tail of the potential. The coefficients $c_3$ and $c_4$ are given by the asymptotic expansion of $_1F_1$. The locations of the zeros of $c_4(\omega_n)$ are the approximate quasinormal modes frequencies.  However,  in general it is not known how  $\tilde{c}_1$, $\tilde{c}_2$ are related to $c_1$, $c_2$ in this approximation. Only in the large mass regime, where $\phi$ is an ingoing wave everywhere, can we fix $\tilde{c}_2 \overset{m \gg 1}{=} c_2$ and $\tilde{c}_1  \overset{m \gg 1}{=}0$. With angular momentum $l$, regime IV will be altered since there is an extra angular momentum barrier. One has to go to a larger mass, depending on $l$, such that the angular momentum barrier also vanishes. However the discussion below is general at large enough $r$.

Finally we consider the limit $\bar{k} \to 0$.  In this limit the function ${}_1F_1$ reduces to ${}_0F_1$ which is related to a Bessel function. Therefore the $r \gg1$ solution (\ref{eqn:heuncapprox})  in the large $m$, $\bar{k} = 0$ regime reduces to 
\begin{align}
  R(r) & \overset{ \bar{k} =0}{=}  \tilde{c}_1 \Gamma(1 +\Delta_{m,l})m^{-\Delta_{m,l}}  \sqrt{r }  J_{\Delta_{m,l}} ( 2m \sqrt{r-1})  \nonumber \\
  &\quad   + \tilde{ c}_2 \Gamma(1 -\Delta_{m,l})m^{\Delta_{m,l}}  \sqrt{r }  J_{-\Delta_{m,l}} ( 2m \sqrt{r-1})  ,
\end{align}
where we have used the following identity, 
\begin{align*}
&  \quad \lim_{\bar{k} \to 0} {}_1F_1 \left( i \bar{k} + i \frac{m^2}{2 \bar{k}} +  \frac{1 \pm \Delta_{\omega,l}}{2},1 \pm \Delta_{\omega,l}, 2 i \bar{k} (r-1) \right) \\
&= \Gamma(1 \pm \Delta_{m,l}) m^{\pm \Delta_{m,l}} (r-1)^{ \pm \frac{1}{2} \Delta_{m,l}} J_{ \pm \Delta_{m,l}} ( 2m \sqrt{r-1} ) .
\end{align*}
If $m^2(r-1)$ is large the asymptotic expansion of the Bessel function gives
\begin{equation}
R(r) \sim r^{\frac{1}{4}} \cos\big( 2m\sqrt{r-1} + {\rm phase}\big) .
\end{equation}
Therefore for $\bar{k} = 0$ {\it the field profile $\phi(r)$ drops as $r^{-3/4}$ for all $l$ } in the region $ r \gg \frac{1}{m^2 r_s} $. The discussion about regime IV in Sec. \ref{sec:schwarzschild} is always valid while other regimes may behave differently for non-zero $l$. 

\section{Asymptotics for $l = \bar{k} = 0$} \label{appendix:asymptotics}
In appendix \ref{app:KG} we obtained solutions to the wave equation in terms of confluent Heun functions.  Here we study these solutions in more detail and obtain approximations valid
in various regimes.  After some preliminary definitions we specialize to fields with no angular momentum ($l = 0$) and frequency $\omega = m$ (equivalently $\bar{k} = 0$).  For related discussion
see \cite{Fiziev:2009}.

The confluent Heun function, ${\sf HeunC}(\alpha,\beta,\gamma,\delta,\eta;z)$, is defined by the following differential equation\footnote{Note that in the literature e.g. in \cite{MR1392976} the Heun function is written in terms of
a different set of parameters, related to those appearing in \eqref{eqn:heundef}, \eqref{eqn:heunbc} by
\begin{align}
 &``\epsilon" = \alpha,\quad  ``\gamma" = 1+ \beta,\quad ``\delta" = 1 + \gamma ,  \nonumber \\
& ``\alpha" = \frac{1}{2}\alpha(2+ \beta + \gamma  ), \quad ``q" =\frac{1}{2} \left( -\beta(1+ \gamma) + \alpha (1+ \beta) - 2 \eta -\gamma \right).
\end{align}
}
\begin{multline} \label{eqn:heundef}
 {\sf HeunC}''(z) + \left(\alpha + \frac{1+\beta}{z} + \frac{1+\gamma}{z-1} \right)  {\sf HeunC}'(z)  \\
 + \left( \frac{(\alpha(2+\beta+ \gamma )+ 2\delta) z + \beta(1+\gamma)  -\alpha (1+\beta) + 2\eta +\gamma }{2z(z-1)} \right)  {\sf HeunC}(z) =0,
\end{multline}
with boundary conditions
\begin{equation}
\label{eqn:heunbc}
 {\sf HeunC}(0) =1 ,\quad {\sf HeunC}'(0) =  \frac{ \beta(1+\gamma)  -\alpha (1+\beta) + 2\eta +\gamma}{2(1+\beta)}.
\end{equation}
 Its power series expansion around $z=0$, ${\sf HeunC}(z) = \sum_{n=0}^{\infty} a_n z^n$ for $|z|<1$, obeys the following recurrence relation, 
\begin{align}
  &P_n a_n  = Q_n a_{n-1} +  R_n a_{n-2}, \nonumber \\
  & P_n = n(n + \beta), \quad Q_n = (n-1)( n+\beta + \gamma - \alpha) + \frac{\beta(1+\gamma)  -\alpha (1+\beta) + 2\eta + \gamma }{2} , \nonumber \\
   & R_n = (n-2)\alpha + \frac{1}{2}\alpha(\beta + \gamma +2 ) + \delta.
\end{align} 
Carrying out this expansion for ${\sf HeunC}(0,\beta,0,\delta,-\delta,z)$, gathering terms to second order in $\beta$ and first order in $\delta$, and then re-summing the $z$-expansion we find that\footnote{Here ${\rm dilog}\,(1-\chi) = \sum_{n=1}^\infty {\chi^n \over n^2}$.}
\begin{eqnarray}
{\sf HeunC}(0,\beta,0,\delta,-\delta,z) &=& 1 - {\beta \over 2} \log (1 - z) + {\beta^2 \over 4} \Big(\log(1 - z) - {\rm dilog}\,(1-z)\Big)\nonumber \\
&& \quad - {\delta \over 2} \Big(z- \log(1 - z)\Big) + \cdots
\end{eqnarray}
The Heun function of interest appears in (\ref{k=0}).  It involves the case $\beta = -2im$, $\delta = -m^2$, $z = 1 - r$ for which this expansion reads
\begin{equation}
\label{HCsmall}
{\sf HeunC} = 1 + i m \log r - {1 \over 2} m^2 \Big(r + 3 \log r - 2 \, {\rm dilog} \, r - 1\Big) + {\cal O}\left(m^3\right) \, ,
\end{equation}

Before proceeding let us examine the expansion in powers of $m$ more closely.  The result (\ref{HCsmall}) is valid up to quadratic order in the small mass $m$.  Strictly speaking this expansion is valid
as $m \rightarrow 0$ with all other parameters fixed.  But other parameters in the problem -- in particular the radius $r$ -- may be large, which means it is important to understand the nature of the expansion a little better.
We start with the first non-trivial term.  Restoring the Schwarzschild radius, and treating the $\log$ as ${\cal O}(1)$, the first non-trivial term is ${\cal O}(mr_s)$.  It will be the dominant correction for sufficiently small mass
(regime I in the paper).  At second order the most important term for large radius $\sim m^2 r$.  Restoring the Schwarzschild radius we see that, in addition to $mr_s$, another expansion parameter in the problem is $m\sqrt{rr_s}$.
The second-order term should really be thought of as ${\cal O}\big((m \sqrt{rr_s})^2\big)$.  Even when the $m$ expansion is valid, the second-order term can dominate over the first-order term if the radius is large enough (this is
regime II in the paper).  Finally, when $m\sqrt{r r_s} > 1$ the expansion in powers of $m$ breaks down.  This is regimes III and IV in the paper.

To develop approximations valid in regimes III and IV let us return to the differential equation (\ref{radial}), which for $l = \bar{k} = 0$ reads
\begin{equation}
\label{Radialk=0}
\left(-\partial_{r_*}^2 - {m^2 \over r} + {1 \over r^3} - {1 \over r^4}\right) R(r) = 0 .
\end{equation}
First let us assume $mr_s < 1$ (regime III).  Then we can divide the radial coordinate into two regions.  In the ``near-field'' region ($r_s < r < 1/m$) the Newtonian potential $-m^2/r$ can be neglected compared to the relativistic corrections
and the differential equation reduces to
\begin{equation}
\left(-\partial_{r_*}^2 + {1 \over r^3} - {1 \over r^4}\right) R(r) = 0 .
\end{equation}
Rather than solve this equation, it is simpler to note that in the near-field region the expansion in powers of $m$ is valid and to leading
order (\ref{HCsmall}) simply gives
\begin{equation}
{\sf HeunC} \approx 1
\end{equation}
(an approximation that is actually valid out to $r \sim 1/m^2$).  From (\ref{k=0}) this corresponds to
\begin{equation}
\label{NearField}
R \approx c_2 r e^{-im \log (r-1)} .
\end{equation}
In the far-field region $1/m < r < \infty$, on the other hand, the Newtonian potential dominates and the differential equation reduces to
\begin{equation}
\left(-\partial_{r_*}^2 - {m^2 \over r} \right) R(r) = 0 .
\end{equation}
This has a solution in terms of Bessel functions,
\begin{eqnarray}
\label{FarField}
R(r) &\approx&\, \tilde{c}_1 \left(J_{2im}\big(2m\sqrt{r-1}\big) + i \sqrt{r-1}\,J_{1+2im}\big(2m\sqrt{r-1}\big)\right) \\
\nonumber
&&+ \,  \tilde{c}_2 \left(J_{-2im}\big(2m\sqrt{r-1}\big) - i \sqrt{r-1}\,J_{1-2im}\big(2m\sqrt{r-1}\big)\right) .
\end{eqnarray}
It is straightforward to match the near- and far-field solutions (\ref{NearField}), (\ref{FarField}) at $mr \sim 1$.  Requiring that the solutions and their radial derivatives are continuous gives, to leading order for small $mr_s$
and assuming real $c_2$,
\begin{equation}
\tilde{c}_1 = \tilde{c}_2^* = {c_2 \over 2 i m} .
\end{equation}
For $r < 1/m^2$ we adopt the approximation (\ref{NearField}), while for $r > 1/m^2$ we expand the Bessel functions in (\ref{FarField}) for large argument.  This leads to
\begin{equation}
\label{RIII}
R(r) \approx \left\lbrace\begin{array}{ll}
c_2 r  e^{-im \log (r-1)} & \qquad 1 < r \ll 1/m^2 \\
{c_2 r^{1/4} \over \sqrt{\pi} m^{3/2}} \cos\left(2m\sqrt{r} - {3 \pi \over 4}\right) & \qquad r \gg 1/m^2
\end{array}\right.
\end{equation}
One could recast this result as an approximation for the Heun function, but this step is not necessary for our purposes.  It is simpler to directly use $\phi(t,r) = e^{-imt} {1 \over r} R(r)$.

Finally, to study the behavior for $mr_s > 1$ (regime IV) we can proceed as follows.  Note that when $mr_s > 1$ the Newtonian potential in (\ref{Radialk=0}) is dominant all the way down to the horizon at $r = 1$.  This means
the far-field solution (\ref{FarField}) is valid all the way to the horizon, so we can impose the near-horizon in-going boundary condition (\ref{NearHorizon0}) directly on the Bessel solution (\ref{FarField}).  This picks out a particular
linear combination of Bessel functions and leads to
\begin{multline}
 {\sf HeunC}(0,-2im,0,-m^2,m^2,1-r) \approx \Gamma(1 - 2im) {1 \over r} \Big(m \sqrt{r-1}\Big)^{2im} \\
  \cdot\left(J_{-2im}\Big(2m\sqrt{r-1}\Big) - i \sqrt{r-1} J_{1 - 2im}\Big(2m\sqrt{r-1}\Big)\right) \, .
 \label{relation}
\end{multline}

\begin{figure}
\begin{center}
\includegraphics[width=9cm]{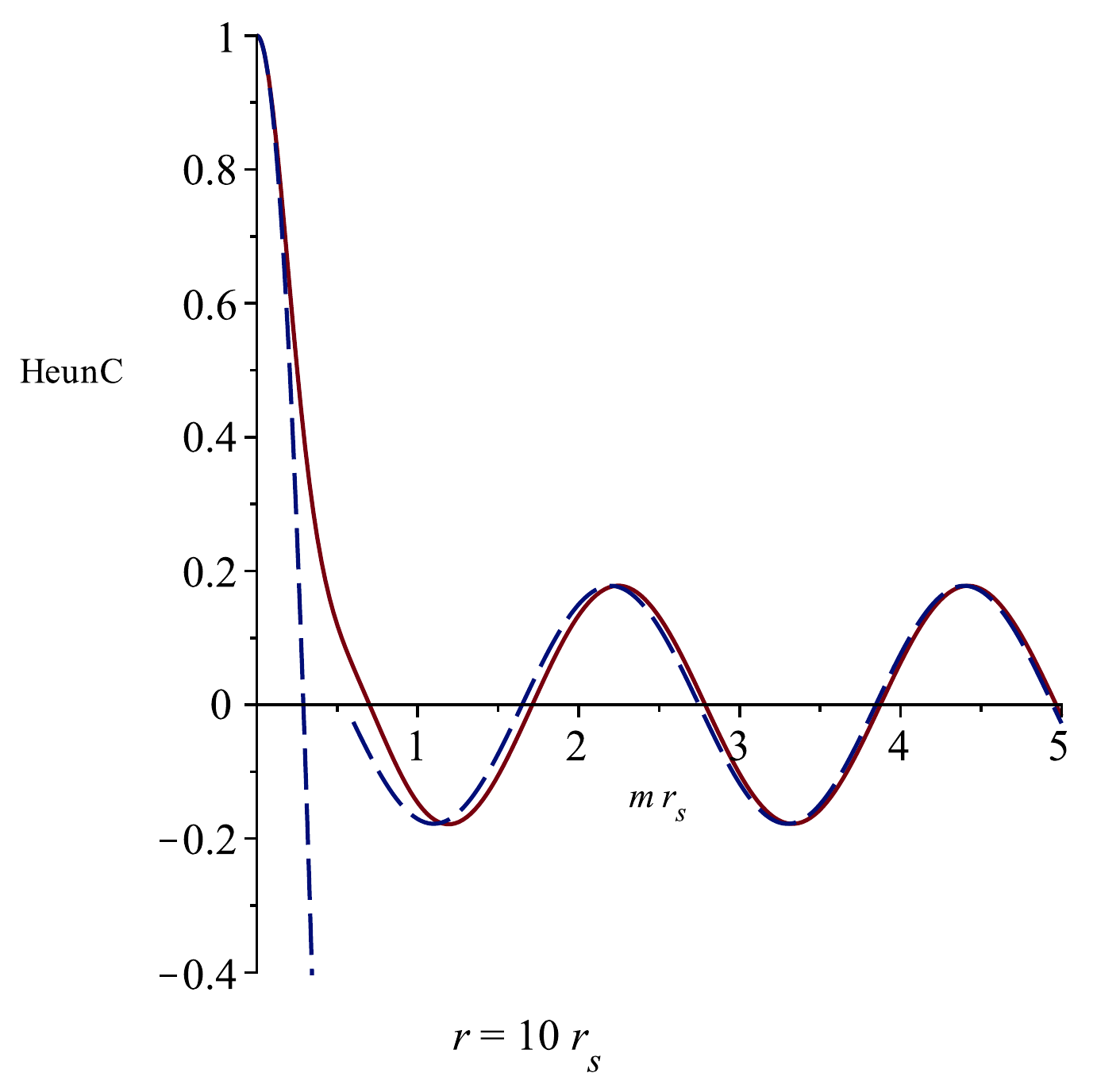}
\end{center}
\caption{Real part of ${\sf HeunC}$ as a function of $m$ for $r = 10$ (solid line).  The approximations \eqref{HCsmall} and \eqref{HCbig} are shown as dashed curves.\label{fig:Heun}}
\end{figure}

To study the right hand side for large $m$ we use\footnote{This is known as approximation by tangents.  It is valid for $\nu \rightarrow \infty$ with $\beta$ fixed.}
\begin{equation}
J_\nu\Big({\nu \over \cos \beta}\Big) = \sqrt{2 \over \pi \nu \tan \beta} \left\lbrace\cos\Big(\nu \tan \beta - \nu \beta - {\pi \over 4}\Big) + {\cal O}(1/\nu)\right\rbrace \, .
\end{equation}
Keeping the leading large-$m$ behavior in \eqref{relation} and expanding the amplitude (not the phase)  for large $r$  leads to the following approximation to the confluent Heun function
\begin{equation}
\label{HCbig}
{\sf HeunC} \approx {1 \over r^{3/4}} e^{-2im\Big(\sqrt{r} - \log(\sqrt{r} + 1) + \log 2 - 1\Big)} .
\end{equation}
The approximations \eqref{HCsmall} and \eqref{HCbig} are illustrated in Fig. \ref{fig:Heun}.

\section{Energy density and flux}
\label{app:flux}
In this section we obtain an expression for the rate at which energy is accreted by the black hole.
We begin by considering a complex scalar field $\phi$ of mass $m$ with action
\begin{equation}
S = \int \D^4x \, \sqrt{-g} \left(-g^{\mu\nu} \partial_\mu \phi^* \partial_\nu \phi - m^2 \phi^*\phi\right)
\end{equation}
and stress tensor
\begin{equation}
T_{\mu\nu} = \partial_\mu \phi \partial_\nu \phi^* + \partial_\mu \phi^* \partial_\nu \phi + g_{\mu \nu} {\cal L} \, .
\label{tmunu}
\end{equation}
Given the background geometry \eqref{staticapp} and assuming a spherically symmetric ansatz for the scalar field, the non-vanishing components of the stress tensor are
\begin{eqnarray}
&& T^{tt} = {1 \over f^2} \left( \vert \partial_t \phi \vert^2 + fg \vert \partial_r \phi \vert^2 + m^2 f \vert \phi \vert^2 \right) \, , \\
&& T^{rt} = - {g \over f} \left(\partial_t \phi \partial_r \phi^* + \partial_t \phi^* \partial_r \phi\right) \, .
\end{eqnarray}
The conservation equation for the stress tensor reads
\begin{equation}
\nabla_\mu T^{\mu\nu} = {1 \over \sqrt{-g}} \partial_\mu (\sqrt{-g} T^{\mu\nu}) + \Gamma^\nu_{\mu\lambda}T^{\mu\lambda} = 0 \, .
\end{equation}
Then, setting  $\nu = t$,  we can massage the previous equation  using $\Gamma^t_{tr} = \Gamma^t_{rt} = {1 \over 2 f} {\D f \over \D r}$ to get
\begin{equation}
\partial_t T^{tt} + {1 \over r^2} \Big({g \over f}\Big)^{1/2} \partial_r (r^2 \Big({f \over g}\Big)^{1/2} T^{rt}) + {1 \over f} {\D f \over \D r} T^{rt} = 0 \, .
\end{equation}
Multiplying this by $r^2 f^{3/2} / g^{1/2}$ leads to the ordinary conservation law
\begin{equation}
\partial_t \left(r^2 {f^{3/2} \over g^{1/2}} T^{tt}\right) + \partial_r \left(r^2 {f^{3/2} \over g^{1/2}} T^{rt} \right) = 0 \, .
\end{equation}
Finally, integrating over a spherical shell we find
\begin{equation}
\D E = 4 \pi r^2 {f^{3/2} \over g^{1/2}} T^{tt} \D r = (\hbox{\rm energy between $r$ and $r + \D r$}) \, .
\end{equation}
Therefore, we can  identify the energy flux (energy per unit time flowing inward) across a sphere of radius $r$ as
\begin{equation}
\Phi = - 4 \pi r^2 {f^{3/2} \over g^{1/2}} T^{rt} \, .
\label{fluxphiapp}
\end{equation}
Let us specialize to solutions that have definite frequency with respect to Killing time $t$ and set
\begin{equation}
\phi(t,r) = e^{-i \omega t} {1 \over r} R(r) \, .
\end{equation}
Then, the energy density becomes independent of time, while the energy flux is independent of both time and radius.  
In fact, the energy flux takes a simple form in terms of the tortoise coordinate $\D r_* = \D r / (fg)^{1/2}$:
\begin{equation}
\label{Flux}
\Phi = 4 \pi i \omega \left(R^* \partial_{r_*} R - R \partial_{r_*} R^*\right) \, ,
\end{equation}
namely, it is proportional to the Wronskian, $W[R^*,R] = R^* \partial_{r_*}R - R \partial_{r_*}R^*$.  Alternatively, it can also be thought of as being  proportional to the probability flux in the effective Schr\"odinger equation \eqref{radial}.  From either point of view $\Phi$ is independent of $r$ as required by energy conservation.  This means that we can evaluate the net energy flux using the near-horizon solution \eqref{NearHorizon},
\begin{equation}
\label{HorizonFlux}
\Phi = 8 \pi \omega^2 \vert c_2 e^{-i \bar{k}}\vert^2 \, ,
\end{equation}
where we retained the factor $e^{-i \bar{k}}$ to account for the possibility of imaginary $\bar{k}$.

The stress tensor for a real scalar field can be obtained simply by taking $\phi$ to be real an inserting a factor of $1/2$.  Thus for a real field
\begin{eqnarray}
\label{Treal}
&& T^{tt} = {1 \over 2f^2} \left((\partial_t \phi)^2 + fg (\partial_r \phi)^2 + m^2 f \phi^2 \right) \, , \\
\nonumber
&& T^{rt} = - {g \over f} \partial_t \phi \partial_r \phi \,.
\end{eqnarray}

\providecommand{\href}[2]{#2}\begingroup\raggedright\endgroup

\end{document}